# Multiphase buffering by ammonia sustains sulfate production in atmospheric aerosols

Guangjie Zheng[1,6], Hang Su[2*], Meinrat O. Andreae[3,4], Ulrich Pöschl[5], Yafang Cheng[1*]

[1] Minerva Research Group, Max Planck Institute for Chemistry, Mainz 55128, Germany.

[2] Institute of Atmospheric Physics, Chinese Academy of Sciences, Beijing, 100029, China

[3] Max Planck Institute for Chemistry, Mainz 55128, Germany

[4] Scripps Institution of Oceanography, University of California San Diego, La Jolla, CA 92037, USA

[5] Multiphase Chemistry Department, Max Planck Institute for Chemistry, Mainz 55128, Germany.

[6] State Key Joint Laboratory of Environmental Simulation and Pollution Control, School of Environment, Tsinghua University, Beijing 100084, China

Corresponding author: Y. Cheng (yafang.cheng@mpic.de); H. Su (suhang@mail.iap.ac.cn).

**Key Points:**

- A characteristic buffering time is proposed to evaluate the competing effects of multiphase buffering and acidification on aerosol pH.

- For most areas, the buffer effect can overwhelm acidification and sustain sulfate production from high pH-favored multiphase reactions.

**Abstract**

Multiphase oxidation of sulfur dioxide ($SO_2$) is an important source of sulfate in the atmosphere. There are, however, concerns that protons produced during $SO_2$ oxidation may cause rapid acidification of aerosol water and thereby quickly shut down the fast reactions favored at high pH. Here, we show that the sustainability of sulfate production is controlled by the competing effects of multiphase buffering and acidification, which can be well described by a characteristic buffering time, $\tau_{buff}$. We find that globally, $\tau_{buff}$ is long enough (days) to sustain sulfate production over most populated regions, where the acidification of aerosol water is counteracted by the strong buffering effect of $NH_4^+/NH_3$. Our results highlight the importance of anthropogenic ammonia emissions and pervasive human influences in shaping the chemical environment of the atmosphere.

**Plain Language Summary**

Aerosol acidity largely regulates the chemistry of atmospheric particles and their environmental effects. Understanding the role of proton ($H^+$) generating reactions is thus a major challenge in understanding and modeling of atmospheric multiphase chemistry. Here, by accounting for the competing effects of multiphase buffering and acidification, we develop a new method to evaluate the role of these reactions on aerosol pH and pH-sensitive reactions. We find that over most populated regions, the buffering effect is so strong that it overwhelms the influence of acidification and sustains sulfate production from high pH-favored multiphase reactions. Our results highlight the importance of anthropogenic ammonia emissions in buffering the acidity of aerosol particles and in regulating the change of atmospheric compositions via multiphase reactions.

## 1. Introduction

Multiphase reactions are important sources of secondary aerosols and fine particulate matter in the atmosphere, influencing air quality, climate, and human health (Akimoto & Hirokawa, 2020; Seinfeld & Pandis, 2016; Su, Cheng, & Pöschl, 2020; Guangjie Zheng et al., 2020; G. J. Zheng et al., 2015). Acidity is a key parameter in multiphase reactions, affecting gas-liquid partitioning and reaction rates (Jang, Czoschke, Lee, & Kamens, 2002; Pye et al., 2020; Tilgner et al., 2021; Guangjie Zheng et al., 2020). The impact of acidity is particularly strong in sulfate formation (Cheng et al., 2016; Seinfeld & Pandis, 2016), a major component of fine particulate matter in the atmosphere (Snider et al., 2016). Elevated aerosol pH of above 4.5 favors aqueous sulfate formation via $SO_2$-$O_3$ and $SO_2$-$NO_2$ pathways, while low pH favors the reaction via $SO_2$-$O_2$ catalyzed by transition metal ions (TMI) (Cheng et al., 2016; Seinfeld & Pandis, 2016). The major oxidation pathways may vary, with $SO_2$-$NO_2$, $SO_2$-$O_3$, $SO_2$-TMI, and $SO_2$-$H_2O_2$ reactions alternatively dominant in different regions depending on the prevailing aerosol pH and precursor concentrations (Keene et al., 1998; Tao et al., 2020).

Because the formation of sulfate is always accompanied by proton ($H^+$) production, there is an arising concern whether the fast reactions favored at high pH can maintain their reaction rates upon acidification (Alexander et al., 2005; Angle et al., 2021; Keene et al., 1998; Laskin et al., 2003; Liu, Chan, & Abbatt, 2021; Pye et al., 2020; Tilgner et al., 2021; W. Wang et al., 2021). The time scale of acidification is thus important in predicting the total amount of sulfate produced and its environmental and climate effects. For example, changing the acidification time scale from ~1 hour to ~1 day would increase the contribution of sea salt chemistry to the global sulfate burden from 1% to 13% (Alexander et al., 2005; Angle et al., 2021; Chameides & Stelson, 1992; Laskin et al., 2003; Liao, Seinfeld, Adams, & Mickley, 2004; Salter et al., 2016). In the polluted boundary layer, a recent comprehensive review on aerosol acidity and multiphase chemistry (Tilgner et al., 2021) suggested that the acidification can be so efficient that haze particles with initial pH above ~4 will be acidified to as low as pH ~1 within 10 seconds. However, Zheng et al.(Guangjie Zheng et al., 2020) found that ammonia could buffer aerosol pH in large areas of the continent, implying a slow acidification process. Observations also show that with fast sulfate production under polluted conditions in the North China Plain, aerosol pH can be maintained at > 4-6 for hours to days depending on meteorological conditions (Ding et al., 2019; Shi et al., 2017). So far, it is still not clear how efficiently the buffer effect can compete with acidification and sustain the pH-sensitive reactions in sulfate production.

## 2. Materials and Methods

### 2.1 Calculating the temporal evolution of aerosol pH upon acidification

**Simulation of the actual buffered aerosols.** To elucidate the temporal evolution of aerosol pH upon acid production by sulfate-forming reactions in polluted regions, we take as an example the severe winter haze conditions in the North China Plain (referred to the NCP scenario hereinafter) (Cheng et al., 2016; Tilgner et al., 2021; Guangjie Zheng et al., 2020). Briefly, sulfate production works like adding sulfuric acid to the system, namely:

$$n_{acid} = 2\,\Delta[SO_4^{2-}] = 2\int P_{SO4}\,dt \tag{1}$$

where $n_{acid}$ is the amount of equivalent strong monoacid added to the system, $\Delta[SO_4^{2-}]$ is the total generated sulfate, and $P_{SO4}$ is the sulfate production rate. The $P_{SO4}$ is a function of aerosol pH, and the aerosol pH is a function of total sulfate. The dependence of $P_{SO4}$ on pH is calculated explicitly as parameterized in Cheng et al. (2016) (Fig. S1), while the influence of ionic strength (Liu & Abbatt, 2021; Liu et al., 2021; Liu, Clegg, & Abbatt, 2020) and other potential mechanisms (Chen et al., 2022; W. Wang et al., 2021) etc. is discussed in SI sect. S1 and Figs. S2-S3. The pH under different total sulfate concentrations can be estimated by the ISORROPIA model (Fountoukis & Nenes, 2007). Therefore, the aerosol pH change due to sulfate formation can be calculated iteratively by accounting for the newly formed sulfate in the ISORROPIA inputs, as briefly detailed below.

The initial conditions (i.e., at reaction time $t = 0$) of aerosol compositions followed the NCP scenario in Zheng et al. (2020) (Table S1), which are kept constant except the sulfate concentrations in the following calculations. The initial pH is estimated by the ISORROPIA model under this condition. Afterwards, the sulfate and pH evolutions are calculated iteratively with a time step $\Delta t$. Briefly, assume at time point $t$ the aerosol pH is pH($t$) and sulfate concentration is $SO_4^{2-}(t)$, the newly formed sulfate during the next time step is then $\Delta SO_4^{2-} = P_{SO4}(pH(t)) \Delta t$. The new sulfate is $SO_4^{2-}(t + \Delta t) = SO_4^{2-}(t) + \Delta SO_4^{2-}$, while pH($t + \Delta t$) is estimated by ISORROPIA with the new sulfate concentration of $SO_4^{2-}(t + \Delta t)$.

During the calculation, we found that the time interval $\Delta t$ can be important, as a too large $\Delta t$ can lead to over-predictions of sulfates formed within one time step, and thus underestimation in the pH evolutions (SI S1). Here, we applied a $\Delta t$ of 0.001 s, which is validated to prevent sulfate over-predictions (SI S1).

**Simulation of the assumed non-buffered aerosols.** The pH evolution of the non-buffered aerosols is calculated with a similar iterative method as the buffered system, with the only difference being the pH~$SO_4^{2-}$ relationships. For non-buffered system, we assume that the newly formed sulfate $\Delta SO_4^{2-}$ works like adding sulfuric acid into the system, and thus $[H^+](t + \Delta t) = [H^+](t) + 2\Delta SO_4^{2-}$, namely pH($t + \Delta t$) = -lg($10^{-pH(t)} + 2\Delta SO_4^{2-}$).

## 2.2 Estimating the pH changes due to acidification within 10 seconds

Another way to evaluate susceptibility of system pH to acidification reactions is to compare the pH changes after a certain time of reactions, e.g., 10 seconds as in Tilgner et al. (2021). To achieve aerosol systems at different initial pH levels, here we scaled the sulfate concentrations while keeping other conditions the same as the initial setting for NCP scenario (Table S1). In that sense, the different initial pH levels actually correspond to different time points following the simulation in sect. 2.1, or different degrees during the acidification process of the given aerosol system. A maximum pH of 6.3 can be achieved this way when sulfate concentration is near 0 (sect. 2.1). As the pH levels are constrained by the known aerosol compositions and not set arbitrarily as did in Tilgner et al. (2021), higher pH levels are not available in this simulation. The subsequent pH changes in the next 10 seconds are then calculated following the iterative calculation methods as described above.

## 2.3 GEOS-Chem model simulation

The global GEOS-Chem model simulation (v11-01) was conducted for 2016, and is detailed elsewhere (Guangjie Zheng et al., 2020). The resolution is 2.5° longitude × 2° latitude with 47 vertical layers. The meteorology fields are based on the Modern-Era Retrospective analysis for Research and Applications, Version 2 (MERRA-2) reanalysis meteorological data product (Gelaro et al., 2017) that is updated every 1–3 hours. Anthropogenic emissions were based on the Emission Database for Global Atmospheric Research (EDGAR v4.2) inventory for 2012 (http://edgar.jrc.ec.europa.eu/overview.php?v=42), while updated over China domain with the Multi-resolution Emission Inventory for China inventory (MEIC; http://meicmodel.org) (v1.3) for 2016. Moreover, we've included the international ship emissions for $SO_2$ based on the Arctic Research of the Composition of the Troposphere from Aircraft and Satellites (ARCTAS) inventory (Eyring, Köhler, Lauer, & Lemper, 2005; Eyring, Köhler, van Aardenne, & Lauer, 2005), and that of CO and NOx based on the International Comprehensive Ocean-Atmosphere Data Set (ICOADS) inventory (C. Wang, Corbett, & Firestone, 2008), and the aircraft NOx emissions from the Aviation Emissions Inventory Code (AEIC v2.1) inventory (Simone, Stettler, & Barrett, 2013; Marc E. J. Stettler, Boies, Petzold, & Barrett, 2013; M. E. J. Stettler, Eastham, & Barrett, 2011). Natural source emissions considered include the biomass burning emissions from Global Fire Emissions Database (GFED v4) (van der Werf et al., 2010), biogenic VOCs and NO emissions calculated by the Model of Emissions of Gases and Aerosols from Nature (MEGAN v2.1) (Guenther et al., 2012), the soil and lightning NOx emissions (Hudman et al., 2012; Sauvage et al., 2007), and the volcanic $SO_2$ emissions from AeroCom point source data (Carn, Yang, Prata, & Krotkov, 2015). While usually run with lower concentration levels, the GEOS-Chem model has been shown to well-reproduce the general spatial and seasonal global distributions of total $NH_3$, $PM_{2.5}$, and thus the aerosol water contents (Luan & Jaeglé, 2013; G. Luo, Yu, & Moch, 2020; M. Luo et al., 2015; Sayer, Thomas, Palmer, & Grainger, 2010; Shephard et al., 2011; Whitburn et al., 2016; Zeng, Tian, & Pan, 2018). See detailed model validations in Wang *et al.* (S. Wang et al., 2020).

## 3. Acidification versus multiphase buffering

Figure 1 shows the simulated acidification process under the NCP scenario upon multiphase oxidation of $SO_2$ (sect. 2.1). The acidification rate turns out to be extremely slow, with pH dropping only by 1 unit over ~9 days of acidification (Fig. 1A), comparable with the lifetime of atmospheric fine particles (Seinfeld & Pandis, 2016). Here, we make a conservative assumption with a constant total $NH_3+NH_4^+$, while with fast replenishment of $NH_3$ (e.g., a constant $NH_3(g)$ as suggested in Weber et al. (Weber, Guo, Russell, & Nenes, 2016)), the acidification could be even slower (Fig. S4 and SI sect. S2). Note that the cases shown in Fig. 1 and Fig. S4 here are to demonstrate the effect of acid production on sulfate-forming reactions. Simultaneous change of other species will confuse this effect and was thus turned off for the sake of easy understanding. In reality, all chemical species were subject to emissions, depositions, meteorology and transport processes, etc. (see discussions on Fig. 3 below).

As suggested by Zheng et al. (Guangjie Zheng et al., 2020), aerosol pH during severe winter haze conditions in the NCP can be largely buffered by the multiphase buffer agent $NH_4^+/NH_3$. To examine the role of the buffer effect in this slow pH drop, we further calculated the multiphase buffer capacity $\beta$ as (Guangjie Zheng et al., 2020; G. Zheng, Su, Wang, Pozzer, & Cheng, 2022):

$$\beta = 2.303\left(\frac{K_w}{[H^+(aq)]} + [H^+(aq)] + \sum_i \frac{K_{a,i}^*[H^+(aq)]}{(K_{a,i}^*+[H^+(aq)])^2}[X_i]_{tot}^*\right), \quad \text{for buffered multiphase systems} \quad (2a)$$

where $K_w$ is water dissociation constant, $K_{a,i}^*$ and $[X_i]_{tot}^*$ represents the effective acid dissociation constant and total equivalent molality of the buffering agent $X_i$, respectively. The $\beta$ is at its local maximum when pH equals $pK_{a,i}^*$. Note that for non-buffered aerosol systems, $X_i = 0$ and $\beta$ is:

$$\beta = 2.303(K_w/[H^+(aq)] + [H^+(aq)]), \quad \text{for non-buffered multiphase systems} \quad (2b)$$

where the remaining terms represent the water self-buffering effect (Guangjie Zheng et al., 2020).

As shown in Fig. 1A, the slow acidification in the first ~9 days is attributed to the buffering effect of ammonia, which has a peak buffer pH (i.e., $pK_{a,i}^*$) of 5.2, and thus buffers efficiently in the pH range of 5.2 ± 1 (Fig. 1A, orange shaded area). For the NCP aerosol system, the initial aerosol pH of 5.4 is close to $pK_{a,i}^*$. As the reactions proceed, the accumulated $n_{acid}$ slowly consumes $\beta$ (Fig. 1B) and decreases pH. After ~9 days' acidification, nearly all the $\beta$ from ammonia is consumed and pH dropped to the edge of the ammonia buffer range (~4.2). Afterwards, in the absence of efficient buffering, pH dropped sharply by 3.7 units (from 4.2 to 0.5) within ~9 hours (the vertical grey shaded area in Figs. 1A to C). When 99% of the ammonia $\beta$ is consumed (i.e., essentially the non-buffered pH ranges of this system), the acidification rate can be particularly high, with pH dropping from 2.9 to 1.3 within 10 minutes. The enhanced $P_{SO4}$ by the TMI-catalyzed pathway at this pH range also accelerated this breaking-out process (Fig. 1C). Especially, because $P_{SO4}$ by the TMI pathway increases with acidity when pH > 2.5, the acidification processes could accelerate itself as it proceeds. When the aerosols are acidified to pH < 2.5, however, $P_{SO4}$ is again suppressed by increasing acidity due to the complex pH-dependence of the $SO_2$-TMI pathway (Fig. S1), and the acidification tends to quench itself. At the low pH range of < 0.5, the acidification is slow again, due to both the enhanced water self-buffering effect (Figs. 1A, B) and the lower $P_{SO4}$ (Fig. 1C).

The characteristics of the acidification process discussed above contrast sharply with that estimated for a non-buffered system (black line vs. blue line in Fig. 2A; sect. 2.2), where the acidification is much quicker. For a wide initial aerosol pH range of 1.5 to 7, the system is acidified to a relatively stable pH level of ~1.6 (1.4-1.7) within 10 seconds. Note that our estimation for the non-buffered system (blue line in Fig. 2A) is different from that of Tilgner et al. (Tilgner et al., 2021) (dashed grey line in Fig. 2A), even after considering the influence of the different $P_{SO4}$ configurations (SI sect. S3). The large pH drop with increasing initial pH level when it is over ~4.5 of Tilgner et al. (Tilgner et al., 2021) seems unrealistic, as the aerosol pH should be a monotonically decreasing function with reaction time during the acidification process for a given system (Fig. 1A; see more discussions in SI sects. S3 and S3). A pattern like the one in Tilgner et al. (Tilgner et al., 2021) is observed only when an inappropriately large time step $dt$ is applied in the iterative calculations, e.g., when $dt$ equals their total reaction time of 10 s (i.e., only one time step is simulated; Fig. S5). Such a large time step will lead to strong overestimation of the acidification and the unrealistic pH response to the initial pH reported by Tilgner et al. (Tilgner et al., 2021) (SI sect. S3).

## 4. Reaction sustainability and characteristic buffering time

The importance of the multiphase buffer effect in stabilizing aerosol acidity is well illustrated by the sharp contrast in acidification processes with versus without the ammonia-buffer effect (Fig. 1A and Fig. 2A). To describe quantitatively the efficiency of the buffer effect and the time scales of buffering against acidification, we introduce a characteristic buffering time, $\tau_{buff}$, which is defined as the reciprocal of the pH change rate as $\tau_{buff} = |d\text{pH}/dt|^{-1}$. According to the definition of buffer capacity $\beta$ (Guangjie Zheng et al., 2020), i.e., the ratio between the amount of acid or base added to the system ($n_{acid}$ or $n_{base}$) and the corresponding pH change ($\beta = - dn_{acid} / d\text{pH}$), it is clear that the efficiency of the multiphase buffer effect against acidification is intrinsically connected with the buffer capacity. We can thus derive the acidification rate ($d\text{pH}/dt$) and $\tau_{buff}$ as:

$$d\text{pH}/dt = (-dn_{acid}/\beta)/dt = (-dn_{acid}/dt)/\beta = -2P_{SO4}/\beta \quad (3)$$

$$\tau_{buff} = |d\text{pH}/dt|^{-1} = 0.5\,\beta/P_{SO4} \quad (4)$$

While in deriving Eqs. 3 and 4 only the sulfate acidifying effect is considered, the above analysis can be easily extended to more general acidifying processes as (see detailed deduction in SI sect. S4.1):

$$\tau_{buff} = |d\text{pH}/dt|^{-1} = \beta / \sum_i v_i P_{acid,i} \quad (5)$$

where $P_{acid,i}$ is the production rate of acidic species $i$, and $v_i$ is the stoichiometric number of $H^+$ associated with the corresponding acid. For example, for sulfate-generating reactions, the corresponding acid is $H_2SO_4$ and v is 2; while v is 1 for chloride-generating reactions.

Here, $\tau_{buff}$ represents the sustainability of the buffering effect (including the water self-buffer effect) for a given system against the acidifying reactions. $\tau_{buff}$ is a property of the aerosol system, which is determined by aerosol and gases compositions through Eqs. (2) and (4). When $\tau_{buff}$ is much larger than the reaction time $t_{rct}$, the buffering effect dominates over the acidifying effect, and the aerosol pH can be sustained. When $\tau_{buff} \ll t_{rct}$, the acidifying effect overwhelms the buffering effect, and a quick pH drop is expected. Take the NCP scenario with a reaction time $t_{rct} = 10$ s for illustration (Fig. 2A). For the buffered system, $\tau_{buff} \gg t_{rct}$ (operationally defined as $\tau_{buff} > 10\,t_{rct}$ here) when pH < 1.6 or pH > 3.2 (Fig. 2B, black line), and therefore the aerosol pH changes little ($\Delta$pH < 0.1) within these two ranges (Fig. 2A, black line). In comparison, for the non-buffered system, $\tau_{buff} \gg t_{rct}$ is satisfied only when pH < 1.6, and the acidification is obvious ($\Delta$pH > 0.1) for any initial pH level above this value. The detailed principles and the quantified relationship of how $\tau_{\text{buff}}$ can be used to characterize the time needed to reach a certain $\Delta$pH, or to predict the $\Delta$pH (or final pH) after a certain reaction time $t_{\text{rct}}$, are discussed in SI sect. S4 and Figs. S6-S7.

## 5. Global distribution of buffering time scale

The above analysis of $\tau_{\text{buff}}$ under the NCP scenario can be extended to other ammonia-buffered atmospheric aerosol systems, based on ambient observations and global simulations. To address the wide concern that whether the acidification due to multiphase reactions would quench the high-pH favored reactions (Angle et al., 2021; Chen et al., 2022; Tilgner et al., 2021; W. Wang et al., 2021), we calculated $\tau_{\text{buff}}$ for the multiphase reactions exclusively (Fig. 3A, Fig. 4A). Figure 3A shows the $\tau_{\text{buff}}$ of sulfate formation under different ambient conditions, characteristic for the Southeastern U.S.A. (SE-US), the North China Plain (NCP), northern India (NI), and western Europe (WE) (see scenario settings in Table S1) (Guangjie Zheng et al., 2020). Under the conditions observed in these regions, the corresponding $\tau_{\text{buff}}$ for the four scenarios are all above 1 day. The shortest $\tau_{\text{buff}}$ is observed for the NCP scenario, considering the higher precursor concentrations and higher reaction rates (Table S1). Figure 4A further compares the global distribution of $\tau_{\text{buff}}$ for the ammonia-buffered regions, with site-specific reactant concentrations and aerosol pH levels based on GEOS-Chem simulations (sect. 2.3). In agreement with the observation-based estimations (Fig. 3A), globally, $\tau_{\text{buff}}$ is generally on the order of 10 days or months, with smaller $\tau_{\text{buff}}$ found in the most polluted regions like northern China and India.

Acidic species generated by gas-phase reactions, however, can also condense onto the particles, contributing to the acidification process. Figure 3B and Figure 4B further consider the influence of the major gas phase $H_2SO_4$ formation pathway by $SO_2$ and OH radical reactions (Cheng et al., 2016; Seinfeld & Pandis, 2016). Based on ambient observations (Fig. 3A, Fig. 4A), this influence is extremely large for the cleaner scenarios of SE-US and WE, reducing $\tau_{buff}$ by 1~2 orders of magnitude. In comparison, the influence on the more polluted scenarios of NI and NCP is much lower, reducing $\tau_{buff}$ only by 3.9 and 1.4 times, respectively. This confirms that globally, gas-phase oxidation of $SO_2$ is one of the most important sources of sulfate, while multiphase reactions play an important role under polluted haze conditions with high aerosol concentrations and high precursor concentrations, such as the NCP and NI scenarios (Fig. S8) (Cheng et al., 2016; Guangjie Zheng et al., 2020; G. J. Zheng et al., 2015). However, even considering the gas phase chemistry, the $\tau_{buff}$ in the buffered regime is still on the scale of days, suggesting the importance of the multiphase buffer effect in resisting extremely quick acidification, which would otherwise have occurred in minutes.

The results based on global simulations also supported this conclusion (Fig. 4B), with $\tau_{buff}$ on scale of days for all multiphase buffered regions. However, it also shows that over most areas, there is a sharp (~2 orders of magnitude) decrease of $\tau_{buff}$ when gas-phase reactions are included (Fig. 4A). This indicates that in terms of the global budget, except for in-cloud production, the sulfate formation in most regions is still dominated by the gas-phase oxidation of $SO_2$, and the multiphase phase production of sulfate in aerosol water will mainly contribute under polluted conditions (Cheng et al., 2016; Tao et al., 2020).

The multiphase buffering effect is important in sustaining a stable aerosol pH and thus a stable reaction rate of pH-sensitive reactions. Depending on the relationship of aerosol acidity and reaction rates, multiphase reactions can be classified into self-amplifying, self-quenching, or weakly-dependent. In a non-buffered system, the self-amplification would proceed quickly, and the self-quenching reactions will be shut down promptly (Figs. 5A, B). Under buffered conditions, however, the buffer effect can hinder the propagation of self-amplifying reactions, or sustain the self-quenching reactions (Figs. 5C, D).

## 6. Conclusions

Both the $SO_2$-$O_3$ and $SO_2$-$NO_2$ reactions are self-quenching. In a non-buffered marine environment, the self-quenching $SO_2$-$O_3$ reaction on sea salt aerosols can quickly shut down on a scale of minutes (Figs. 5A, B) (Alexander et al., 2005; Angle et al., 2021; Seinfeld & Pandis, 2016). However, in ammonia-buffered polluted continental regions and marine regions receiving $NH_3$ from continental sources (Fig. 4), the self-quenching sulfate formation reactions of $SO_2$-$O_3$ and $SO_2$-$NO_2$ can be sustained over days, contributing significantly to sulfate production during haze formation (Fig. 5C) (Cheng et al., 2016), as illustrated above. Even for the partially self-amplifying $SO_2$-TMI reaction, the overall sulfate yield is higher with multiphase buffering. The $SO_2$-TMI reaction would turn from self-amplifying to self-quenching when pH drops below ~2.5 for the NCP scenario (Fig. 1; Fig. S1). Under non-buffered conditions, this rapid acidification would quickly drop the pH to the turnover point, followed by a rapid quenching (Fig. 1). Therefore, the overall sulfate yield would be smaller than under buffered conditions over a time scale of ~ 1 hour.

Our results show that the ammonia multiphase buffering effect is important in sustaining a stable aerosol pH and enhancing the overall sulfate production, thereby impacting the global sulfate burden and haze formation. The stability of aerosol acidity could also influence the production of other components (SI sect. S5), like secondary organics (Carlton et al., 2010; Franco et al., 2021; Hallquist et al., 2009; Jang et al., 2002; Surratt et al., 2007). Our finding highlights the pervasive influence of anthropogenic ammonia emissions in shaping chemical environments and controlling the dominant reaction pathways in the atmosphere.

**Acknowledgments**

This study is support by Max Planck Society (MPG). Y.C. would like to thank the Minerva Program of MPG. G. Z. acknowledges the National Natural Science Foundation of China (22188102).

**Author Contributions**

Y.C. and H.S. designed and led the study. G.Z., Y.C., and H.S. proposed the concept and performed the research. U.P. and M.O.A. discussed the results. H.S., Y.C., G.Z. wrote the manuscript with input from U.P. and M.O.A.

**Conflict of Interest:** Authors declare no competing interests.

**Data availability:** All data used in the analysis are provided in the SI.

**References**

Akimoto, H., & Hirokawa, J. (2020). *Atmospheric Multiphase Chemistry: Fundamentals of Secondary Aerosol Formation*: John Wiley & Sons.
Alexander, B., Park, R. J., Jacob, D. J., Li, Q. B., Yantosca, R. M., Savarino, J., . . . Thiemens, M. H. (2005). Sulfate formation in sea-salt aerosols: Constraints from oxygen isotopes. *Journal of Geophysical Research: Atmospheres, 110*(D10). doi:https://doi.org/10.1029/2004JD005659

Angle, K. J., Crocker, D. R., Simpson, R. M. C., Mayer, K. J., Garofalo, L. A., Moore, A. N., . . . Grassian, V. H. (2021). Acidity across the interface from the ocean surface to sea spray aerosol. *Proceedings of the National Academy of Sciences, 118*(2), e2018397118. doi:10.1073/pnas.2018397118

Carlton, A. G., Bhave, P. V., Napelenok, S. L., Edney, E. O., Sarwar, G., Pinder, R. W., . . . Houyoux, M. (2010). Model Representation of Secondary Organic Aerosol in CMAQv4.7. *Environmental Science & Technology, 44*(22), 8553-8560. doi:10.1021/es100636q

Carn, S. A., Yang, K., Prata, A. J., & Krotkov, N. A. (2015). Extending the long-term record of volcanic SO2 emissions with the Ozone Mapping and Profiler Suite nadir mapper. *Geophysical Research Letters, 42*(3), 925-932. doi:10.1002/2014gl062437

Chameides, W. L., & Stelson, A. W. (1992). Aqueous-phase chemical processes in deliquescent sea-salt aerosols: A mechanism that couples the atmospheric cycles of S and sea salt. *Journal of Geophysical Research: Atmospheres, 97*(D18), 20565-20580. doi:https://doi.org/10.1029/92JD01923

Chen, Z., Liu, P., Wang, W., Cao, X., Liu, Y.-X., Zhang, Y.-H., & Ge, M. (2022). Rapid Sulfate Formation via Uncatalyzed Autoxidation of Sulfur Dioxide in Aerosol Microdroplets. *Environmental Science & Technology, 56*(12), 7637-7646. doi:10.1021/acs.est.2c00112

Cheng, Y., Zheng, G., Wei, C., Mu, Q., Zheng, B., Wang, Z., . . . Su, H. (2016). Reactive nitrogen chemistry in aerosol water as a source of sulfate during haze events in China. *Science Advances, 2*(12), e1601530. doi:10.1126/sciadv.1601530

Ding, J., Zhao, P., Su, J., Dong, Q., Du, X., & Zhang, Y. (2019). Aerosol pH and its driving factors in Beijing. *Atmos. Chem. Phys., 19*(12), 7939-7954. doi:10.5194/acp-19-7939-2019

Eyring, V., Köhler, H. W., Lauer, A., & Lemper, B. (2005). Emissions from international shipping: 2. Impact of future technologies on scenarios until 2050. *Journal of Geophysical Research: Atmospheres, 110*(D17), D17306. doi:10.1029/2004jd005620

Eyring, V., Köhler, H. W., van Aardenne, J., & Lauer, A. (2005). Emissions from international shipping: 1. The last 50 years. *Journal of Geophysical Research: Atmospheres, 110*(D17), D17305. doi:10.1029/2004jd005619

Fountoukis, C., & Nenes, A. (2007). ISORROPIA II: a computationally efficient thermodynamic equilibrium model for K+-Ca2+-Mg2+-NH4+-Na+-SO42--NO3--Cl--H2O aerosols. *Atmos. Chem. Phys., 7*(17), 4639-4659. doi:10.5194/acp-7-4639-2007

Franco, B., Blumenstock, T., Cho, C., Clarisse, L., Clerbaux, C., Coheur, P. F., . . . Taraborrelli, D. (2021). Ubiquitous atmospheric production of organic acids mediated by cloud droplets. *Nature, 593*(7858), 233-237. doi:10.1038/s41586-021-03462-x

Gelaro, R., McCarty, W., Suárez, M. J., Todling, R., Molod, A., Takacs, L., . . . Zhao, B. (2017). The Modern-Era Retrospective Analysis for Research and Applications, Version 2 (MERRA-2). *Journal of Climate, 30*(14), 5419-5454. doi:10.1175/jcli-d-16-0758.1

Guenther, A. B., Jiang, X., Heald, C. L., Sakulyanontvittaya, T., Duhl, T., Emmons, L. K., & Wang, X. (2012). The Model of Emissions of Gases and Aerosols from Nature version 2.1 (MEGAN2.1): an extended and updated framework for modeling biogenic emissions. *Geosci. Model Dev., 5*(6), 1471-1492. doi:10.5194/gmd-5-1471-2012

Hallquist, M., Wenger, J. C., Baltensperger, U., Rudich, Y., Simpson, D., Claeys, M., . . . Wildt, J. (2009). The formation, properties and impact of secondary organic aerosol: current and emerging issues. *Atmos. Chem. Phys., 9*(14), 5155-5236. doi:10.5194/acp-9-5155-2009

Hudman, R. C., Moore, N. E., Mebust, A. K., Martin, R. V., Russell, A. R., Valin, L. C., & Cohen, R. C. (2012). Steps towards a mechanistic model of global soil nitric oxide emissions: implementation and space based-constraints. *Atmos. Chem. Phys., 12*(16), 7779-7795. doi:10.5194/acp-12-7779-2012

Jang, M., Czoschke, N. M., Lee, S., & Kamens, R. M. (2002). Heterogeneous Atmospheric Aerosol Production by Acid-Catalyzed Particle-Phase Reactions. *Science, 298*(5594), 814-817. doi:10.1126/science.1075798

Keene, W. C., Sander, R., Pszenny, A. A. P., Vogt, R., Crutzen, P. J., & Galloway, J. N. (1998). Aerosol pH in the marine boundary layer: A review and model evaluation. *Journal of Aerosol Science, 29*(3), 339-356. doi:https://doi.org/10.1016/S0021-8502(97)10011-8

Laskin, A., Gaspar, D. J., Wang, W., Hunt, S. W., Cowin, J. P., Colson, S. D., & Finlayson-Pitts, B. J. (2003). Reactions at Interfaces As a Source of Sulfate Formation in Sea-Salt Particles. *Science, 301*(5631), 340-344. doi:10.1126/science.1085374

Liao, H., Seinfeld, J. H., Adams, P. J., & Mickley, L. J. (2004). Global radiative forcing of coupled tropospheric ozone and aerosols in a unified general circulation model. *Journal of Geophysical Research: Atmospheres, 109*(D16). doi:https://doi.org/10.1029/2003JD004456

Liu, T., & Abbatt, J. P. D. (2021). Oxidation of sulfur dioxide by nitrogen dioxide accelerated at the interface of deliquesced aerosol particles. *Nature Chemistry*. doi:10.1038/s41557-021-00777-0

Liu, T., Chan, A. W. H., & Abbatt, J. P. D. (2021). Multiphase Oxidation of Sulfur Dioxide in Aerosol Particles: Implications for Sulfate Formation in Polluted Environments. *Environmental Science & Technology, 55*(8), 4227-4242. doi:10.1021/acs.est.0c06496

Liu, T., Clegg, S. L., & Abbatt, J. P. D. (2020). Fast oxidation of sulfur dioxide by hydrogen peroxide in deliquesced aerosol particles. *Proceedings of the National Academy of Sciences*, 201916401. doi:10.1073/pnas.1916401117

Luan, Y., & Jaeglé, L. (2013). Composite study of aerosol export events from East Asia and North America. *Atmos. Chem. Phys., 13*(3), 1221-1242. doi:10.5194/acp-13-1221-2013

Luo, G., Yu, F., & Moch, J. M. (2020). Further improvement of wet process treatments in GEOS-Chem v12.6.0: Impact on global distributions of aerosol precursors and aerosols. *Geosci. Model Dev. Discuss., 2020*, 1-39. doi:10.5194/gmd-2020-11

Luo, M., Shephard, M. W., Cady-Pereira, K. E., Henze, D. K., Zhu, L., Bash, J. O., . . . Jones, M. R. (2015). Satellite observations of tropospheric ammonia and carbon monoxide: Global distributions, regional correlations and comparisons to model simulations. *Atmospheric Environment, 106*, 262-277. doi:https://doi.org/10.1016/j.atmosenv.2015.02.007

Pye, H. O. T., Nenes, A., Alexander, B., Ault, A. P., Barth, M. C., Clegg, S. L., . . . Zuend, A. (2020). The acidity of atmospheric particles and clouds. *Atmos. Chem. Phys., 20*(8), 4809-4888. doi:10.5194/acp-20-4809-2020

Salter, M. E., Hamacher-Barth, E., Leck, C., Werner, J., Johnson, C. M., Riipinen, I., . . . Zieger, P. (2016). Calcium enrichment in sea spray aerosol particles. *Geophysical Research Letters, 43*(15), 8277-8285. doi:https://doi.org/10.1002/2016GL070275

Sauvage, B., Martin, R. V., van Donkelaar, A., Liu, X., Chance, K., Jaeglé, L., . . . Fu, T. M. (2007). Remote sensed and in situ constraints on processes affecting tropical tropospheric ozone. *Atmos. Chem. Phys., 7*(3), 815-838. doi:10.5194/acp-7-815-2007

Sayer, A. M., Thomas, G. E., Palmer, P. I., & Grainger, R. G. (2010). Some implications of sampling choices on comparisons between satellite and model aerosol optical depth fields. *Atmos. Chem. Phys., 10*(22), 10705-10716. doi:10.5194/acp-10-10705-2010

Seinfeld, J. H., & Pandis, S. N. (2016). *Atmospheric chemistry and physics: from air pollution to climate change*: John Wiley & Sons.

Shephard, M. W., Cady-Pereira, K. E., Luo, M., Henze, D. K., Pinder, R. W., Walker, J. T., . . . Clarisse, L. (2011). TES ammonia retrieval strategy and global observations of the spatial and seasonal variability of ammonia. *Atmos. Chem. Phys., 11*(20), 10743-10763. doi:10.5194/acp-11-10743-2011

Shi, G., Xu, J., Peng, X., Xiao, Z., Chen, K., Tian, Y., . . . Russell, A. G. (2017). pH of Aerosols in a Polluted Atmosphere: Source Contributions to Highly Acidic Aerosol. *Environmental Science & Technology, 51*(8), 4289-4296. doi:10.1021/acs.est.6b05736

Simone, N. W., Stettler, M. E. J., & Barrett, S. R. H. (2013). Rapid estimation of global civil aviation emissions with uncertainty quantification. *Transportation Research Part D: Transport and Environment, 25*, 33-41. doi:https://doi.org/10.1016/j.trd.2013.07.001

Snider, G., Weagle, C. L., Murdymootoo, K. K., Ring, A., Ritchie, Y., Stone, E., . . . Martin, R. V. (2016). Variation in global chemical composition of PM2.5: emerging results from SPARTAN. *Atmos. Chem. Phys., 16*(15), 9629-9653. doi:10.5194/acp-16-9629-2016

Stettler, M. E. J., Boies, A. M., Petzold, A., & Barrett, S. R. H. (2013). Global Civil Aviation Black Carbon Emissions. *Environmental Science & Technology, 47*(18), 10397-10404. doi:10.1021/es401356v

Stettler, M. E. J., Eastham, S., & Barrett, S. R. H. (2011). Air quality and public health impacts of UK airports. Part I: Emissions. *Atmospheric Environment, 45*(31), 5415-5424. doi:https://doi.org/10.1016/j.atmosenv.2011.07.012

Su, H., Cheng, Y., & Pöschl, U. (2020). New Multiphase Chemical Processes Influencing Atmospheric Aerosols, Air Quality, and Climate in the Anthropocene. *Accounts of Chemical Research, 53*(10), 2034-2043. doi:10.1021/acs.accounts.0c00246

Surratt, J. D., Lewandowski, M., Offenberg, J. H., Jaoui, M., Kleindienst, T. E., Edney, E. O., & Seinfeld, J. H. (2007). Effect of Acidity on Secondary Organic Aerosol Formation from Isoprene. *Environmental Science & Technology, 41*(15), 5363-5369. doi:10.1021/es0704176

Tao, W., Su, H., Zheng, G., Wang, J., Wei, C., Liu, L., . . . Cheng, Y. (2020). Aerosol pH and chemical regimes of sulfate formation in aerosol water during winter haze in the North China Plain. *Atmos. Chem. Phys., 20*(20), 11729-11746. doi:10.5194/acp-20-11729-2020

Tilgner, A., Schaefer, T., Alexander, B., Barth, M., Collett Jr, J. L., Fahey, K. M., . . . McNeill, V. F. (2021). Acidity and the multiphase chemistry of atmospheric aqueous particles and clouds. *Atmos. Chem. Phys., 21*(17), 13483-13536. doi:10.5194/acp-21-13483-2021

van der Werf, G. R., Randerson, J. T., Giglio, L., Collatz, G. J., Mu, M., Kasibhatla, P. S., . . . van Leeuwen, T. T. (2010). Global fire emissions and the contribution of deforestation, savanna, forest, agricultural, and peat fires (1997–2009). *Atmos. Chem. Phys., 10*(23), 11707-11735. doi:10.5194/acp-10-11707-2010

Wang, C., Corbett, J. J., & Firestone, J. (2008). Improving Spatial Representation of Global Ship Emissions Inventories. *Environmental Science & Technology, 42*(1), 193-199. doi:10.1021/es0700799

Wang, S., Su, H., Chen, C., Tao, W., Streets, D. G., Lu, Z., . . . Cheng, Y. (2020). Natural gas shortages during the “coal-to-gas” transition in China have caused a large redistribution of air pollution in winter 2017. *Proceedings of the National Academy of Sciences, 117*(49), 31018-31025. doi:doi:10.1073/pnas.2007513117

Wang, W., Liu, M., Wang, T., Song, Y., Zhou, L., Cao, J., . . . Ge, M. (2021). Sulfate formation is dominated by manganese-catalyzed oxidation of SO2 on aerosol surfaces during haze events. *Nature Communications, 12*(1), 1993. doi:10.1038/s41467-021-22091-6

Weber, R. J., Guo, H., Russell, A. G., & Nenes, A. (2016). High aerosol acidity despite declining atmospheric sulfate concentrations over the past 15 years. *Nature Geosci, 9*(4), 282-285. doi:10.1038/ngeo2665

http://www.nature.com/ngeo/journal/v9/n4/abs/ngeo2665.html#supplementary-information

Whitburn, S., Van Damme, M., Clarisse, L., Bauduin, S., Heald, C. L., Hadji-Lazaro, J., . . . Coheur, P.-F. (2016). A flexible and robust neural network IASI-NH3 retrieval algorithm. *Journal of Geophysical Research: Atmospheres, 121*(11), 6581-6599. doi:10.1002/2016jd024828

Zeng, Y., Tian, S., & Pan, Y. (2018). Revealing the Sources of Atmospheric Ammonia: a Review. *Current Pollution Reports, 4*(3), 189-197. doi:10.1007/s40726-018-0096-6

Zheng, G., Su, H., Wang, S., Andreae, M. O., Pöschl, U., & Cheng, Y. (2020). Multiphase buffer theory explains contrasts in atmospheric aerosol acidity. *Science, 369*(6509), 1374-1377. doi:10.1126/science.aba3719

Zheng, G., Su, H., Wang, S., Pozzer, A., & Cheng, Y. (2022). Impact of non-ideality on reconstructing spatial and temporal variations in aerosol acidity with multiphase buffer theory. *Atmos. Chem. Phys., 22*(1), 47-63. doi:10.5194/acp-22-47-2022

Zheng, G. J., Duan, F. K., Su, H., Ma, Y. L., Cheng, Y., Zheng, B., . . . He, K. B. (2015). Exploring the severe winter haze in Beijing: the impact of synoptic weather, regional transport and heterogeneous reactions. *Atmos. Chem. Phys., 15*(6), 2969-2983. doi:10.5194/acp-15-2969-2015

**Figures and Tables**

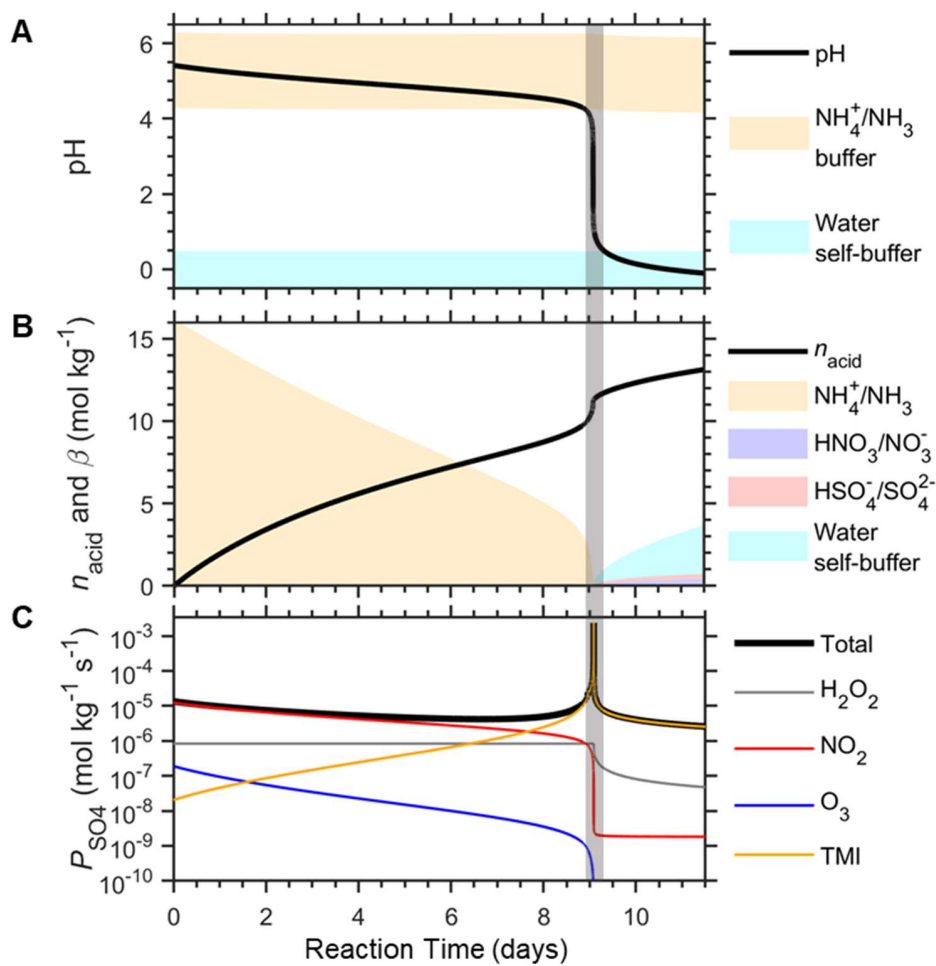

**Figure 1. Acidification related to sulfate formation under severe haze conditions.** Evolution of (**A**) aerosol pH and the buffered pH ranges, (**B**) the amount of equivalent strong monoacid added to the system ($n_{acid}$) due to the produced sulfates, the buffer capacity $\beta$, and (**C**) the sulfate production rate ($P_{SO4}$) along the reaction time. The different buffer agents of $NH_4^+/NH_3$, $HNO_3/NO_3^-$, $HSO_4^-/SO_4^{2-}$ and water self-buffering are marked with color shading in (**A**) and (**B**). The total $P_{SO4}$ shown in (**C**) is the sum of the sulfate formation from S(IV) oxidation by $H_2O_2$, $NO_2$, $O_3$ and TMI pathways. Here, we assume a constant total $NH_3+NH_4^+$ in the gas-particle system.

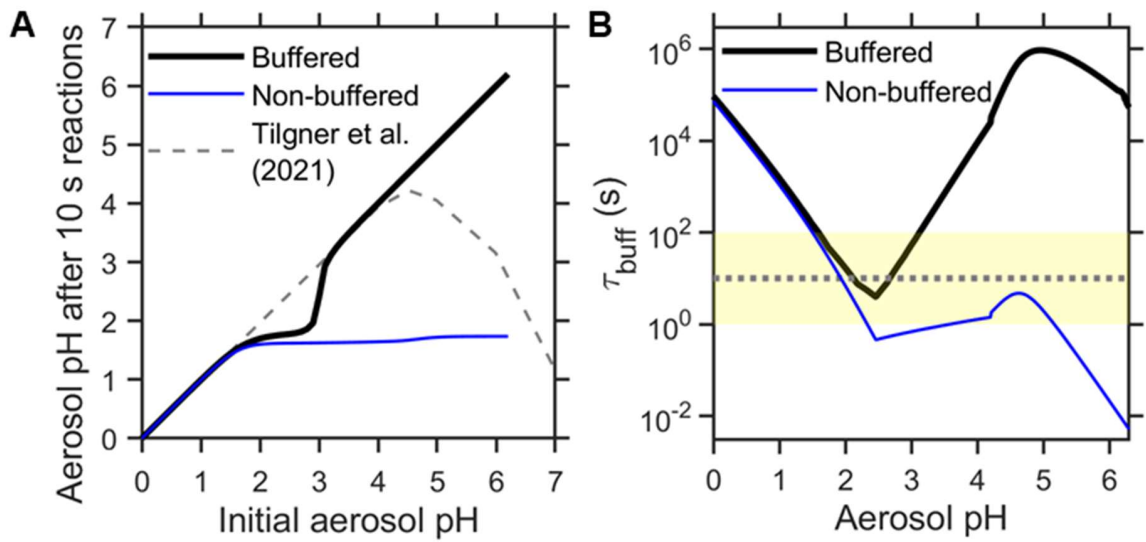

**Figure 2. Acidification characteristics for buffered and non-buffered aerosols under severe haze conditions.** The scenario settings of precursor concentration and production rates are the same as in Fig. 1 (sect. 2.1, "NCP" scenario in Table S1). **(A)** The aerosol pH after 10 seconds (s) for aerosols with different initial pH levels. **(B)** Characteristic buffering time $\tau_{buff}$. The grey dotted line indicates the reaction time $t_{rct}$ of 10 s as used in **(A)**, while the shaded area indicates the range when $\tau_{buff}$ is comparable with the $t_{rct}$ of 10 s (operationally defined as $0.1\ t_{rct} < \tau_{buff} < 10\ t_{rct}$ here).

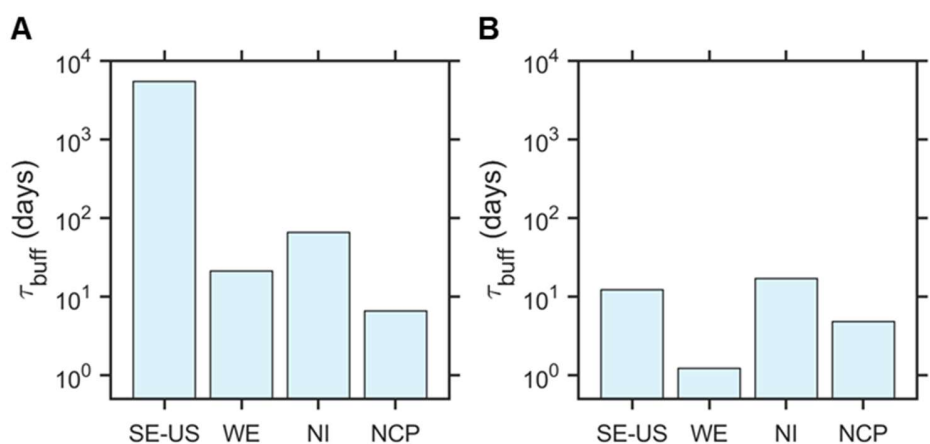

**Figure 3. Observation-based estimations of characteristic buffering time, $\tau_{buff}$, under different ambient conditions.** The scenarios shown are characteristic for the Southeastern U.S.A. (SE-US), western Europe (WE), northern India (NI), and the North China Plain (NCP) (see scenario settings in Table S1). Note that the $\tau_{buff}$ values shown here correspond to different time periods (see Table S1). Only sulfate formation from the multiphase reactions in aerosol waters are considered in (A), while sulfate formation due to the $SO_2$ oxidation in the gas-phase by OH radicals is included in (B).

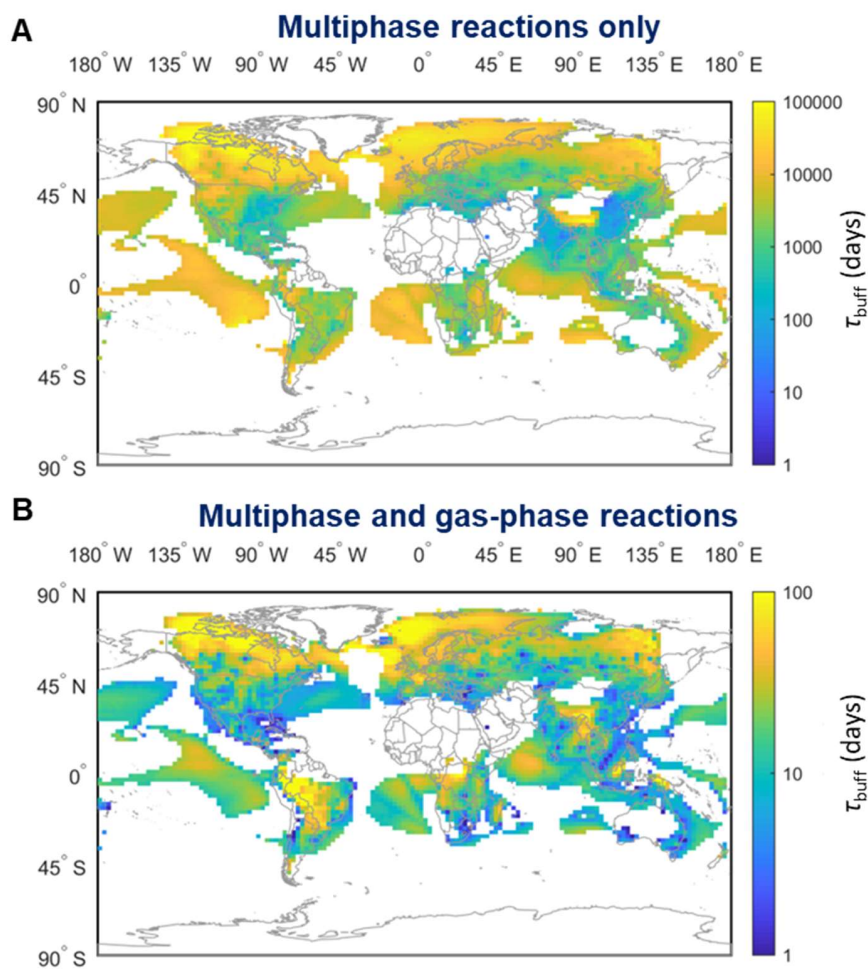

**Figure 4. Global distribution of characteristic buffering time, $\tau_{\text{buff}}$, for ammonia-buffered regions upon sulfate formations.** The $\tau_{\text{buff}}$ is estimated with site-specific reactant concentrations and aerosol pH levels based on GEOS-Chem simulations in 2016 (sect. 2.3). Only sulfate formation from the multiphase reactions in aerosol waters are considered in (A), while sulfate formation due to the $SO_2$ oxidation in the gas-phase by OH radicals is included in (B).

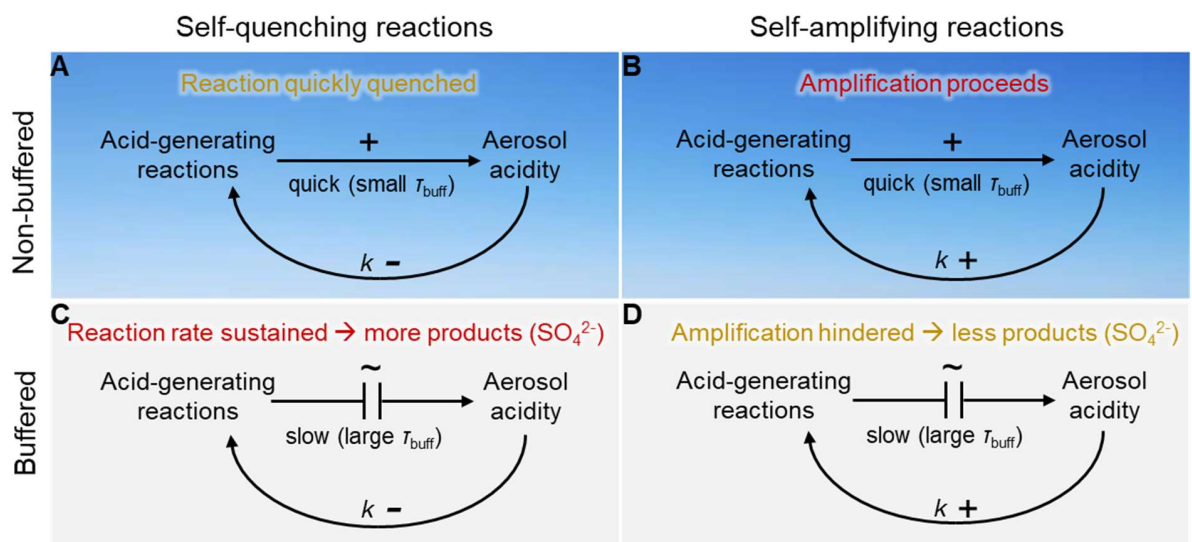

**Figure 5. Influence of acidification versus the multiphase buffer effect on multiphase reactions.** The left panels (**A**, **C**) and the right panels (**B**, **D**) are for self-quenching reactions and self-amplifying reactions, respectively. The upper panels (**A**, **B**) and the lower panels (**C**, **D**) are for non-buffered and buffered systems, respectively. Here, $k$ is the reaction rate, and $\tau_{buff}$ is the characteristic buffering time. The acid-generating reactions will increase aerosol acidity, while this increase is obvious only when the reaction time is comparable with the $\tau_{buff}$. The resultant acidity changes would in turn influence the acid generation rates $k$. In a non-buffered system with small $\tau_{buff}$ (**A**, **B**), such feedback loops are quick, and the acid-generating reactions will be quickly quench or amplified. In a buffered system with large $\tau_{buff}$ (**C**, **D**), however, the acidity adjusts slowly with the generated acids, and the acid-generating reactions can be sustained at stable rates.

Supporting Information for

# Multiphase buffering by ammonia sustains sulfate production in atmospheric aerosols

**Guangjie Zheng[1,6], Hang Su[2,7,*], Meinrat O. Andreae[3,4], Ulrich Pöschl[2], Yafang Cheng[1,5,*]**

[1] Minerva Research Group, Max Planck Institute for Chemistry, Mainz 55128, Germany.

[2] Multiphase Chemistry Department, Max Planck Institute for Chemistry, Mainz 55128, Germany.

[3] Max Planck Institute for Chemistry, Mainz 55128, Germany

[4] Scripps Institution of Oceanography, University of California San Diego, La Jolla, CA 92037, USA

[5] Department of Precision Machinery and Precision Instrumentation, University of Science and Technology of China, Hefei 230026, China

[6] State Key Joint Laboratory of Environmental Simulation and Pollution Control, School of Environment, Tsinghua University, Beijing 100084, China

[7] Institute of Atmospheric Physics, Chinese Academy of Sciences, Beijing, 100029, China

Corresponding author: Y. Cheng (yafang.cheng@mpic.de); H. Su (h.su@mpic.de).

**Contents of this file**

Supplementary Text S1 to S5

Figs. S1 to S8

Table S1

References

**Supplementary Text**

**S1: Influence of $P_{SO4}$ parameterizations**

In this study, we adopted the $P_{SO4}$ parameterization based on Cheng et al. (12), which is well constrained by the state-of-art models. Recent studies have suggested the importance of ionic strength on known sulfate-forming reactions (8, 14, 15) and the potential contribution of other mechanisms (5, 6) on sulfate formation. The applicability of these new updates remains to be fully illustrated and validated with observations and modelling studies. In that circumstance, the framework that we proposed (Eqs. 2 to 4) are still valid and can be easily applied with the updated $P_{SO4}$ parameterization.

Here, we conducted a rough sensitivity study on the potential influence of $P_{SO4}$ parameterization on $\tau_{buff}$, with the assumption that the total multiphase sulfate formation rate is increased by 10 times, i.e. $P_{SO4, new} = 10\ P_{SO4}$. Correspondingly, the $\tau_{buff}$ considering multiphase reactions only would be reduced by 10 times, while it was still well over 1 day under most cases (Fig. S2A, Fig. S3A). The influence would be even smaller when gas-phase reactions are taken into account (Fig. S2B, Fig. S3B). Therefore, the major conclusion of this study that the $\tau_{buff}$ is long enough will still hold even if the $P_{SO4}$ is increased by 10 times.

**S2 Acidification considering the $NH_3$ emissions**

In calculating the acidification processes shown in Fig. 1, we assumed a constant total (gas and particle phase) ammonia, i.e., no $NH_3$ replenishment through emissions. Under ambient conditions, however, a constant $NH_3(g)$ can be assumed with stable emissions (50). Figure S4 shows the acidification processes related to sulfate formation under such conditions. In Fig. S4, the $NH_3(g)$ is kept constant while $NH_4^+$ will increase with the increasing $SO_4^{2-}$, thus the total ammonia is increasing.

As shown in Fig. S4, with the $NH_3$ replenishment through emissions, the acidification would be even slower. The system would stay in the $NH_3$ buffering ranges even after ~ 2 weeks' acidification, with only a slight decrease in pH and $\beta$. The still slightly decreasing $\beta$ with increasing total ammonia is due to that:

$$\beta = 2.303 \left( \frac{K_w}{[H^+(aq)]} + [H^+(aq)] + \sum_i \frac{K_{a,i}^*[H^+(aq)]}{(K_{a,i}^* + [H^+(aq)])^2} [X_i]_{tot}^* \right)$$

and thus the $\beta$ of ammonia is positiviely correlated to the total equivalent particle phase ammonia, [NH$_3$]$_{tot}^*$, in moles per kilogram of water (5). The [NH$_3$]$_{tot}^*$ is proportional to $c_{NH3,\,tot}$ / AWC, where $c_{NH3,\,tot}$ is the total ambient concentration of total ammonia in micromoles per cubic meter of air, and AWC is the aerosol water content in micrograms per cubic meter of air. As stated above, the $\Delta c_{NH3,\,tot}$ equals $\Delta c_{NH4+}$ and is thus proportional to $\Delta SO_4^{2-}$. Meanwhile, when other parameters are kept constant, $\Delta$AWC is also proportional to $\Delta SO_4^{2-}$, and the increasing AWC with formed sulfate tends to decrease $\beta$. That is, both $c_{NH3,\,tot}$ and AWC will increase with sulfate formations, while the net result is a slight decrease in $\beta_{NH3}$.

## S3 Importance of time step *dt* in calculating the acidification processes

The acidification characteristics under severe winter haze conditions in the North China Plain as estimated in our study is completely different from that of Tilgner et al. (6). We note that Tilgner et al. (6) adopted the non-buffered assumption, and applied a different $P_{SO4}$ parameterization from ours. To be directly comparable, we have adopted exactly the same $P_{SO4}$ and non-buffered assumption as in their study, and the remaining difference comes totally from the calculation method. As shown in Fig. S5, however, there's still large difference in the simulated acidification characteristics. While our simulation indicates a monotonically increasing final pH with initial pH (Fig. S5a, grey line), Tilgner et al. estimated a larger pH drop with increasing initial pH levels when pH is over ~4.5 (black dashed line in Fig. S5a).

The decreasing pattern shown in Tilgner et al. (6) is unrealistic, as the aerosol pH should be a monotonically decreasing function with reaction time during the acidification process for a given system (Fig. 1a). A non-monotonic curve like the one in Tilgner et al. (6) is observed in our calculation only when the time step *dt* is too big, where *dt* is the time interval applied in each step during the iterative calculations (sect. 2.1). The simulated pH evolutions are found to be sensitive to the *dt* chosen, especially at higher pH levels (Fig. S5a). When *dt* is set directly to the total simulated reaction time of 10 seconds with only one-step calculation, the calculated final pH would be decreasing for initial pH over ~4.3 (red line in Fig. S5a) due to the rapidly increasing sulfate formation rate (6), which is the most similar result with Tilgner et al. (6). When further shortening the simulation time intervals *dt* and increasing the iteration steps, however, the final pH would be gradually stabilized at ~4.2 for all initial pH levels above. The simulation results converged for *dt* smaller than 0.001s (Fig. S5a).

The infinitesimal *dt* assumption should be applied under actual conditions, as the characteristic time of aqueous-phase dissociation reactions (~2×10$^{-7}$ s)(2) is much smaller than that of chemical reactions. That is, the system pH should decrease instantaneously in response to the formed sulfate, and the $P_{SO4}$ would also adjust instantaneously to the pH changes. Without the time lags in $P_{SO4}$ to pH responses, the time required for a given system to change within given pH levels should be fixed. For illustration, for the system with an initial pH of 5.2, assume it takes $\Delta t_1$ for the aerosol pH to drop from 5.2 to 4.8 and $\Delta t_2$ to drop from 4.8 to 4.4 (blue line in Fig. S5b), then in the given total time of $\Delta t = \Delta t_1 + \Delta t_2$ it would be acidified to 4.4. In comparison, for the aerosol system with initial pH of 4.8, the time required for it to drop to 4.4 is the same as in the previous processes under the same $P_{SO4}$, namely being $\Delta t_2$ (red line in Fig. S5b). In $\Delta t$, however, the final pH for this system should be lower than 4.4 considering the extra acidification time of $\Delta t_1$.

Inappropriately big *dt* would overestimate the acidification and therefore the pH changes. A time step *dt* of 0.001 s is found to be small enough and is applied in our calculation.

## S4 General representation, interpretation and applications of characteristic buffering time, $\tau_{buff}$

### S4.1 General representation of $\tau_{buff}$

Here we consider pH changes caused by redox reactions that generate acidic substances, and the reactions that generate alkaline substances can be derived similarly. Assume the reaction produces acidic substances at the rate of:

$$dn_{acid} / dt = \sum_i v_i P_{acid, i} \quad (S1)$$

where $n_{acid}$ is the amount of equivalent strong monoacids added to the system, $P_{acid,i}$ is the production rate of acidic species *i*, and $v_i$ is the stoichiometric number of H$^+$ associated with the corresponding acid. For example, for sulfate-generating reactions, the corresponding acid is H$_2$SO$_4$ and v is 2; while v is 1 for chloride-generating reactions.

And based on the definition of multiphase buffer capacity $\beta$ of (5):

$$\beta = - dn_{acid} / d\text{pH} \quad (S2)$$

we can derive the acidification rate as:

$$d\text{pH} / dt = (-dn_{acid} / \beta) / dt = -\beta^{-1} \sum_i v_i P_{acid, i} \quad (S3)$$

The temporal evolution of pH can thus be calculated iteratively with known $P_{acid}$ and $\beta$, where $\beta$ can be estimated by Eq. 2 in the main text (5). The characteristic buffering time $\tau_{buff}$ is thus:

$$\tau_{buff} = |d\text{pH}/dt|^{-1} = \beta / \sum_i v_i P_{acid,i} \tag{S4}$$

### S4.2 Interpretation of $\tau_{buff}$

Based on the definition of $\tau_{buff}$ (i.e., $\tau_{buff} = dt/|d\text{pH}|$), we can derive that:

$$dt = \tau_{buff} |d\text{pH}| \tag{S5}$$

That is, within the given reaction time $t_{rct}$, we have:

$$\int_0^{t_{rct}} dt = \int_{\text{pH}_0}^{\text{pH}_1} \tau_{buff} d\text{pH} \tag{S6}$$

And the relationship between reaction time $t_{rct}$ and pH changes is therefore:

$$t_{rct} = \int_{\text{pH}_0}^{\text{pH}_1} \tau_{buff} |d\text{pH}| = \overline{\tau}_{buff} \Delta\text{pH} \tag{S7}$$

where $\Delta\text{pH} = |\text{pH}_1 - \text{pH}_0|$, and $\text{pH}_0$ and $\text{pH}_1$ are the pH before and after $t_{rct}$ of reactions, respectively. That is, in the plot of $\tau_{buff}$ versus pH, the area under $\tau_{buff}$ curves between $\text{pH}_0$ and $\text{pH}_1$ should equal $t_{rct}$. The $\overline{\tau}_{buff}$ represents the average $\tau_{buff}$ in this pH range.

### S4.3 Application of $\tau_{buff}$

The $\tau_{buff} = f(\text{pH})$ curve can be used to quantitatively characterize the acidification time scales, i.e., either the time needed for the aerosol pH to change a certain value (Fig. S6a), or to predict the extent of pH changes after $t_{rct}$ (Fig. S6b). Here we used the simplest case of the $SO_2$-$NO_2$ reaction in the non-buffered system as an illustration, while more complex cases can be explained similarly. This reaction is a self-quenching reaction. That is, as the reaction proceeds, pH tends to decrease (i.e., $\text{pH}_1 < \text{pH}_0$) and thus $P_{SO4}$ tends to decrease.

Figure S6a illustrates conceptually how $\tau_{buff}$ can be used to predict the time needed for the aerosol pH to change by a certain value (here operationally defined as $\Delta\text{pH} = 0.1$). Based on Eq. S3, we see the time needed for a system to change from $\text{pH}_0$ to $\text{pH}_1$ is the integral of $\tau_{buff}(\text{pH})$ curve between $\text{pH}_0$ and $\text{pH}_1$. Assume the average $\tau_{buff}$ in this pH range is $\overline{\tau}_{buff}$, then the time required for $\Delta\text{pH} = 0.1$ is $0.1 \overline{\tau}_{buff}$, and a higher $\overline{\tau}_{buff}$ indicates a longer time required for the same pH changes. Therefore, for the same $\Delta\text{pH}$, the time needed for pH to changes from $A_0$ to $A_1$ (Fig. S6a, yellow shaded area) is much smaller than that of $B_0$ to $B_1$ (Fig. S6a, blue shaded area).

Figure S6b illustrates conceptually how the $\tau_{\text{buff}} = f(\text{pH})$ curve can be used to predict the extent of pH changes after $t_{\text{rct}}$. Based on Eq. S3, when $t_{\text{rct}}$ is fixed, the integral between $\text{pH}_0$ and $\text{pH}_1$ is fixed. At initial pH levels when $\tau_{\text{buff}}$ is comparable with $t_{\text{rct}}$ (e.g., $0.1\ t_{\text{rct}} < \tau_{\text{buff}} (\text{pH}_0) < 10\ t_{\text{rct}}$, point A to $C_0$), the required pH change is sensitive to the shape of the $\tau_{\text{buff}}(\text{pH})$ curve. The $\Delta\text{pH}$ will be larger if $\tau_{\text{buff}}$ near $\text{pH}_0$ is generally low, and changes slower (e.g., $A_0$ to $A_1$, red shaded area) and smaller if $\tau_{\text{buff}}$ nearby is generally higher or increases rapidly (e.g., $B_0$ to $B_1$, blue shaded area). In comparison, at initial pH levels where $\tau_{\text{buff}} \gg t_{\text{rct}}$ (e.g., point C), the height is too large that the required pH change would be small, and $\text{pH}_1 \approx \text{pH}_0$. Especially, if $\tau_{\text{buff}} > 10 t_{\text{rct}}$ is always satisfied at the pH intervals near $\text{pH}_0$, the required pH change would thus satisfy $\Delta\text{pH} < 0.1$ unit (grey shaded area). In contrast, at initial pH intervals where $\tau_{\text{buff}} \ll t_{\text{rct}}$ (e.g., pH levels higher than point $A_0$), the integrated area would be too small until the pH decreased to point $A_0$, so that the final pH at pH levels higher than $A_0$ would be basically the same as if the initial pH is at point $A_0$.

Figure 2b can well explain the acidification rate observed in Fig. 1, with $\tau_{\text{buff}}$ on the order of days when pH is over 4.2 (i.e., ammonia buffered range) or below 1.3 (i.e., water self-buffering pH ranges), but dropped sharply to the order of hours in the middle pH ranges. The $\tau_{\text{buff}}$ is comparable with 10 s in the pH ranges of 1.6~3.2, which can explain the obvious pH changes in 10 s within this range (Fig. 2a, black line).

### *S4.4 Comparison of different $\tau_{\text{buff}}$ definitions.*

The characteristic time is commonly used to describe how fast a process is, while its definition may have different forms. Figure S7 compares other potential definitions of $\tau_{\text{buff}}$ with the current one for the NCP scenario, i.e., the time required for aerosol pH to change 0.1 and 1 unit, and the $[\text{H}^+]$ e-folding time (i.e., when $[\text{H}^+]$ increase by $e$ (Euler's number) times, i.e., pH decrease by $\lg(e) = 0.43$ units). Generally, these potential definitions are on the same order with current definitions.

The different time scales at the same pH are mainly due to the different $\Delta\text{pH}$ (Eq. S3), while the similar scale but apparently shifting in between $\tau_{\text{buff}}$ and the time required for $\Delta\text{pH} = 1$ is due to the difference between $\tau_{\text{buff}}$ and $\overline{\tau}_{\text{buff}}$.

## **S5 Validity of constant pH assumption based on observations**

A constant pH assumption is usually applied in formation rate calculations based on hourly or daily observations, for sulfate and secondary organic compounds alike. However, if the acidification due to sulfate formation is so fast that pH would decrease sharply in seconds as suggested in Tilgner et al. (6), then this assumption could not be applied. As has been discussed extensively above, this is not true due to the multiphase buffering effect.

Large pH variations are possible considering the sensitivity of aerosol pH to meteorology parameters such as RH and temperature (5), which can vary much on time scale of hours. During polluted conditions when meteorology is relatively stable, however, the aerosol pH can be maintained at high levels for hours to days (21, 51).

Overall, the time scale of meteorology variations is much shorter than the estimated $\tau_{\text{buff}}$. That is, the pH changes due to sulfate formations is expected to be much smaller than the variation in meteorology parameters. Therefore, in the current estimation of aerosol pH based on hourly average chemical compositions and meteorology parameters, the influence due to sulfate formation is of minor concern. A constant pH assumption is valid only if the meteorology is stable for the observation hour. The same is true in global simulations.

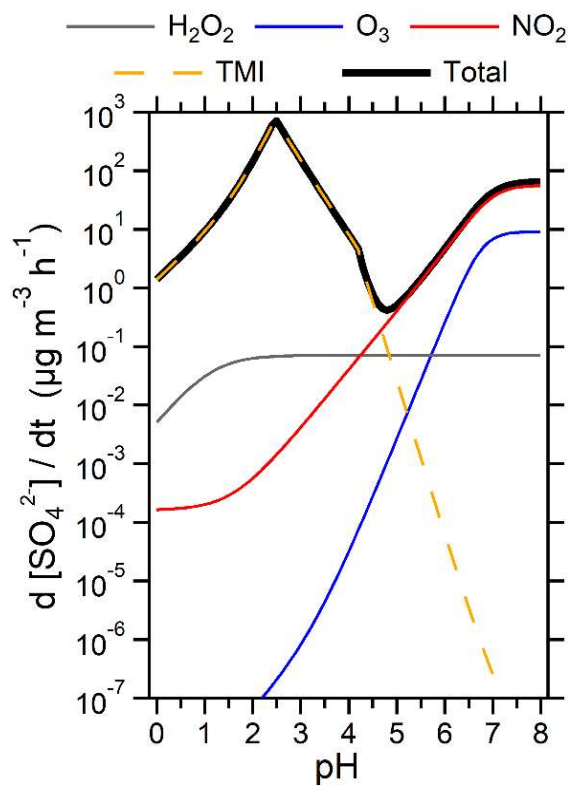

**Fig. S1. Dependence of multiphase sulfate formation on aerosol pH under severe winter haze conditions in the North China Plain.** Here sulfate formation from S(IV) oxidation by $H_2O_2$, $NO_2$, $O_3$ and TMI pathways are considered, and the total is the sum of these pathways. The concentration of fine particulate matter ($PM_{2.5}$) during severe winter haze events is $\geq$ 300 μg m$^{-3}$ (9). The TMI pathway dominated in lower pH ranges of 4.5, with the peak sulfate formation rate of this pathway at pH 2.5. The overall sulfate formation rate, $P_{SO4}$, is assumed to be the sum of these pathways.

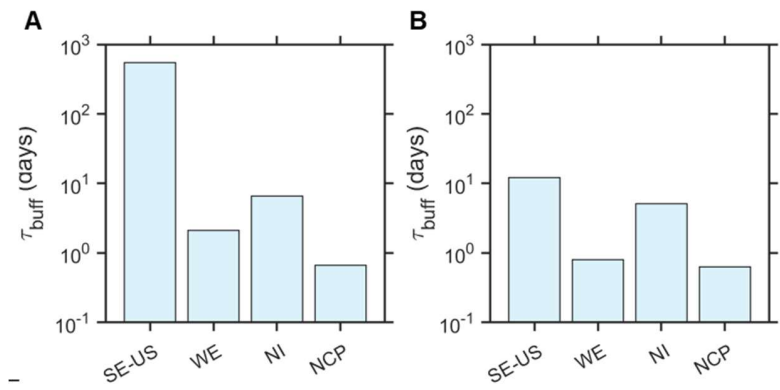

**Fig. S2. Same as Fig. 3, but the multiphase reaction rates are increased by 10 times.**

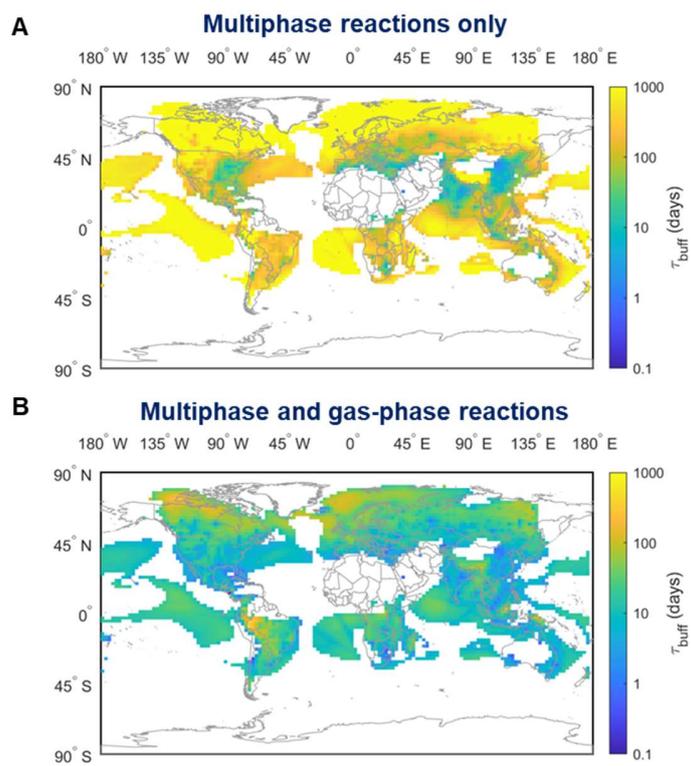

**Fig. S3. Same as Fig. 4, but the multiphase reaction rates are increased by 10 times.**

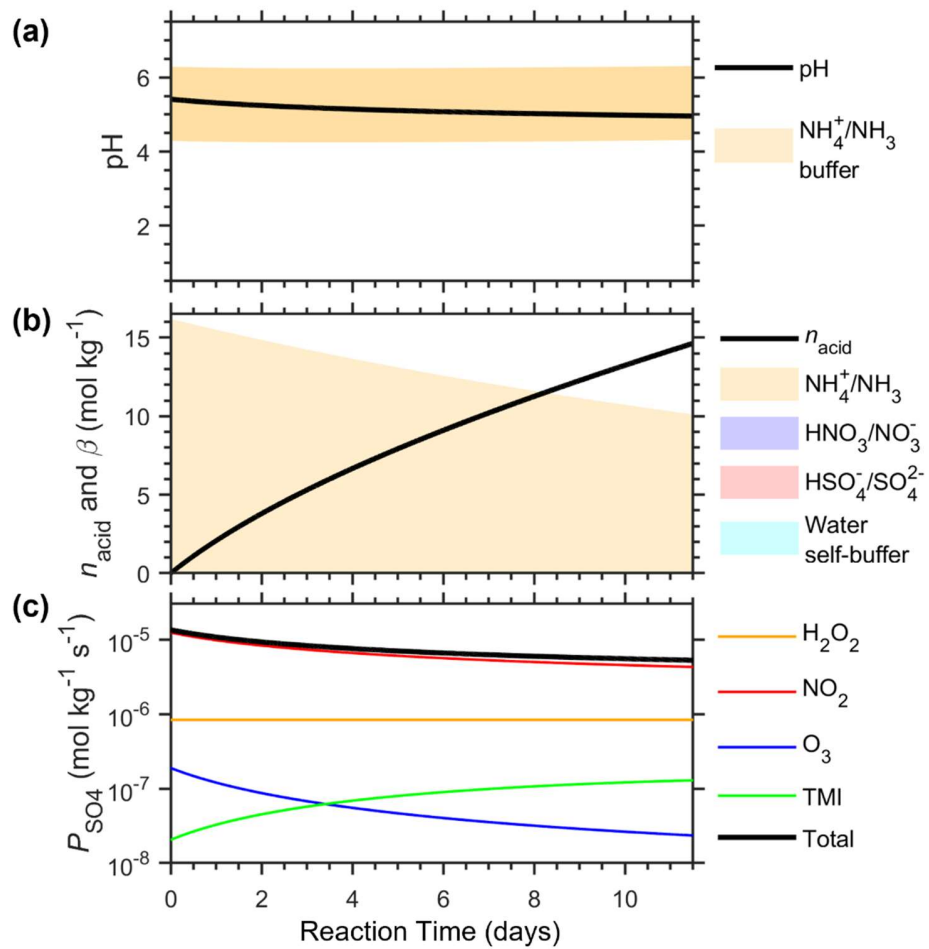

**Fig. S4. Acidification related to sulfate formation under severe haze conditions considering NH₃ replenishment by emissions.** The plots are the same as Fig. 1, but the NH$_3$(g) mixing ratios is assumed to be constant.

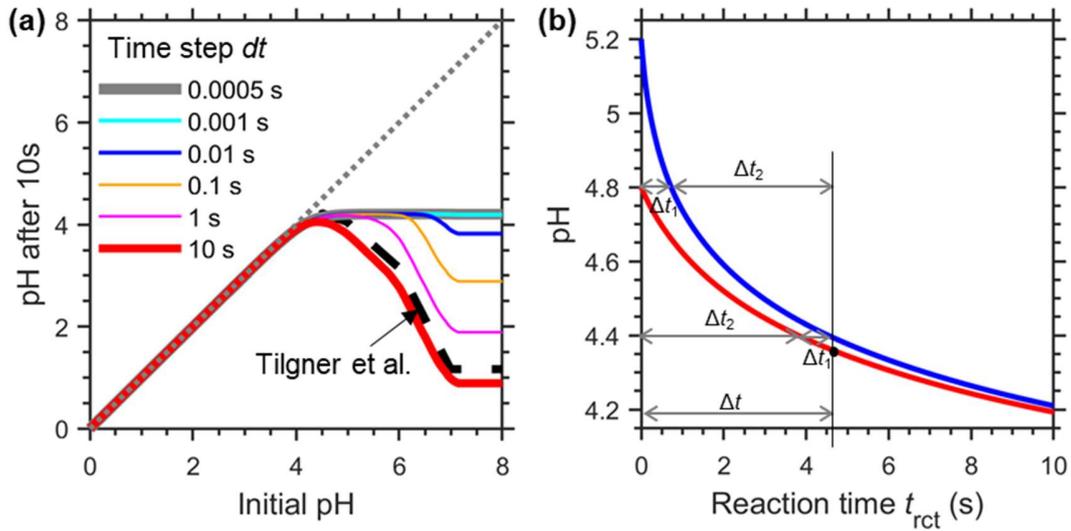

**Fig. S5. Importance of time step *dt* in estimating the pH changes upon acidification from sulfate formation.** (**a**) Dependence of pH after 10 seconds of reaction with initial pH calculated with different *dt* levels, and (**b**) the evolution processes of the system with two initial pH levels, simulated with a small *dt* of 0.001 s. To be directly comparable with the study by Tilgner et al. (6), the system is assumed to be non-buffered as in that study, with the $P_{SO4}$ parameterization following the one shown in Fig. 6 of Tilgner et al.(6). Within a given reaction time of Δ*t*, the system with initial pH of 5.2 (blue line in (b)) will drop to 4.4, while with initial pH of 4.8 (red line in (b)) the final pH is lower than 4.4 (the black dot in (b)). The Δ$t_2$ represents the time required for the system to change from 4.8 to 4.4, which should be the same for a given system with different initial pH levels. Therefore, a non-monotonic curve like Tilgner et al. (6) (black dashed line in (a) is unrealistic, and is observed only when the time step *dt* is inappropriately large of ~10 s (i.e., one-step simulation).

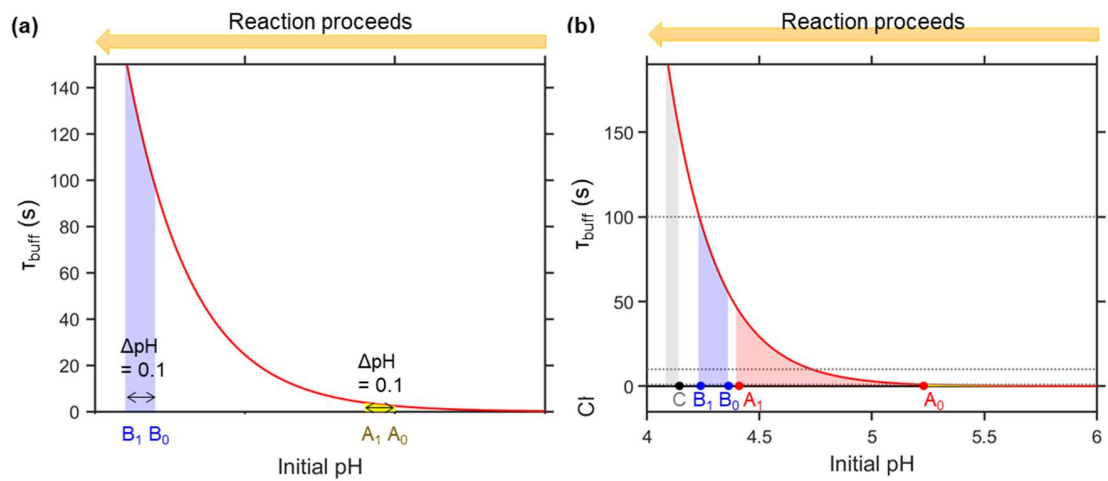

**Fig. S6. Concept and application of the characteristic buffering time $\tau_{buff}$.** (a) The time required for the pH to decrease by 0.1 unit. The integrated area of the shaded regions represents the required reaction time, which is larger with higher $\tau_{buff}$. (b) The pH changes after a given reaction time $t_{rct}$ (assumed to be 10 s here). The dashed grey horizontal lines indicate the $t_{rct}/10$, $t_{rct}$, and $10t_{rct}$, respectively. The red, blue, and grey shaded areas are with the same integrated value of $t_{rct}$ (i.e., 10 s). When the $\tau_{buff}$ is higher, the corresponding pH change is smaller.

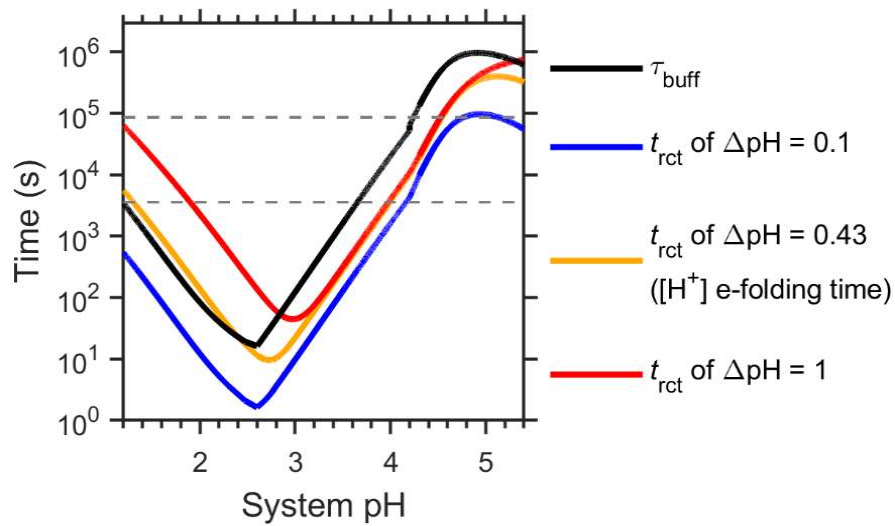

**Fig. S7. Comparison of different definitions of the characteristic buffering time $\tau_{\text{buff}}$ for the NCP scenario** (see Table S1 for scenario settings)**.**

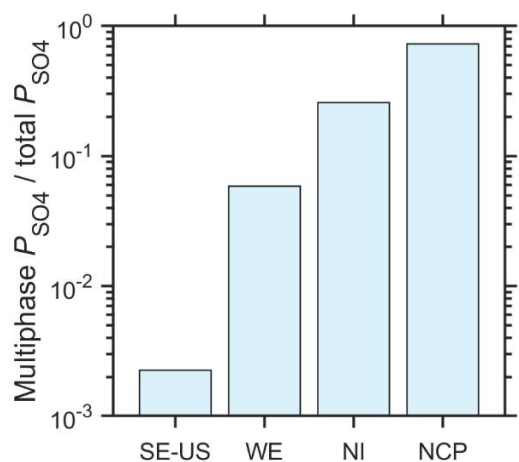

**Fig. S8. Fraction of sulfate formation rate, $P_{SO4}$, by multiphase reactions vs. the total (multiphase + gas-phase reactions of OH radicals).** Here the $P_{SO4}$ is based on ambient observations under conditions characteristic for the Southeastern U.S.A. (SE-US), western Europe (WE), northern India (NI), and the North China Plain (NCP) for ammonia-buffered regions (Table S1). The $P_{SO4}$ fraction can reflect but is not exactly the accumulated sulfate produced through multiphase vs. gas-phase reactions.

**Table S1.**

Scenario settings used in manuscript Figs. 1, 2 and 3(A,C), and supplementary material Fig. S7[a].

| Scenario Name | NCP | NI[b] | WE[c] | SE-US | Related calculation[d] |
|---|---|---|---|---|---|
| $Na^+$ (μg m$^{-3}$) | 5.40 | 1.80 | 1.22 | 0.03 | $β$ |
| $SO_4^{2-}$ (μg m$^{-3}$) | 156.03 | 22.70 | 2.69 | 1.73 | $β$ |
| Total $NH_3$[e] (μg m$^{-3}$) | 168.69 | 26.60 | 8.48 | 0.78 | $β$ |
| Total $HNO_3$ (μg m$^{-3}$) | 84.02 | 12.90 | 3.35 | 0.45 | $β$ |
| Total $HCl$ (μg m$^{-3}$) | 24.01 | 1.70 | 0.12 | 0.02 | $β$ |
| $Ca^{2+}$ (μg m$^{-3}$) | 7.80 | 2.30 | 0.00 | 0.00 | $β$ |
| $K^+$ (μg m$^{-3}$) | 12.60 | 3.20 | 0.14 | 0.00 | $β$ |
| $Mg^{2+}$ (μg m$^{-3}$) | 0.00 | 1.30 | 0.00 | 0.00 | $β$ |
| RH | 0.73 | 0.73 | 0.73 | 0.73 | $β$ |
| Temperature (K) | 269 | 288[f] | 288[f] | 298 | $β$, $P_{SO4}$ |
| $PM_{2.5}$ (μg m$^{-3}$) | 634 | 172 | 21 | 8.9 | / |
| $SO_2$ (ppb) | 40 | 3.9 | 12.6 | 0.3 | $P_{SO4}$ |
| $NO_2$ (ppb) | 66 | 12.4 | 23.6 | 0.6 | $P_{SO4}$ |
| $O_3$ (ppb) | 1 | 46.0[g] | 8.7[g] | 45.7[g] | $P_{SO4}$ |
| $H_2O_2$ (ppb) | 0.01[h] | 1[i] | 1[i] | 1[i] | $P_{SO4}$ |
| Fe (μg m$^{-3}$) | 0.90 | 2.5E-04[g] | 1.2E-05[g] | 6.7E-05[g] | $P_{SO4}$ |
| Mn (μg m$^{-3}$) | 0.08 | 5.5E-05[g] | 2.6E-06[g] | 1.5E-05[g] | $P_{SO4}$ |
| OH (molecules cm$^{-3}$) | 1.0E+6[i] | 1.0E+6[i] | 1.0E+6[i] | 1.0E+6[i] | Gas phase $P_{SO4}$ |
| Reference | (4, 52-55) | (56) | (57) | (58) | / |

[a] Scenario setting are based on observations of the corresponding references unless otherwise noted (shaded cells; see detailed data source below).

[b] Here the NI site is an urban site at Kanpur, India sampled in winter 2007-2008 (56).

[c] Here the WE site is an urban site at Antwerp, Belgium sampled in winter 2002-2003 (57).

[d] These settings are used in the calculation of $β$ (based on ISORROPIA outputs) and /or $P_{SO4}$ (based on the method outlined in Cheng et al. (9). The $PM_{2.5}$ concentration is shown for reference.

[e] Here the total $NH_3$ of NCP is estimated using the gas phase $NH_3$-NOx relationship (9), while all the others are the sum of reported gas and particle phase $NH_3$ concentration.

[f] Here the temperature is set as in-between the NCP and SE-US.

[g] Based on the global simulation of the corresponding grids (sect. 2.3). For $O_3$, the median values of the corresponding months are used. For TMI concentrations, the total Fe is assume to be 3.5% of total dust mass, and Mn is 2% that of total Fe(59, 60). The simulated concentrations are scaled by the ratio of simulated versus observed $PM_{2.5}$ concentrations. The total soluble Fe(III) and Mn(II) fraction is assumed to be 2% and 50% that of total Fe and Mn (9).

[h] Based on the revised simulation of Zheng et al. (61), and is the average when $PM_{2.5}$ is over 300 µg m$^{-3}$.

[i] A constant concentration is assumed to represent daily averages. Note that for the more polluted NCP scenario, the simulated OH radical concentrations is actually much lower (61), and the daily average assumed here should be viewed as an upper limit. The same may apply to the more polluted NI scenario.

**References:**

1. Akimoto H & Hirokawa J (2020) *Atmospheric Multiphase Chemistry: Fundamentals of Secondary Aerosol Formation* (John Wiley & Sons).
2. Seinfeld JH & Pandis SN (2016) *Atmospheric chemistry and physics: from air pollution to climate change* (John Wiley & Sons).
3. Su H, Cheng Y, & Pöschl U (2020) New Multiphase Chemical Processes Influencing Atmospheric Aerosols, Air Quality, and Climate in the Anthropocene. *Accounts of Chemical Research* 53(10):2034-2043.
4. Zheng GJ, *et al.* (2015) Exploring the severe winter haze in Beijing: the impact of synoptic weather, regional transport and heterogeneous reactions. *Atmos. Chem. Phys.* 15(6):2969-2983.
5. Zheng G, *et al.* (2020) Multiphase buffer theory explains contrasts in atmospheric aerosol acidity. *Science* 369(6509):1374-1377.
6. Tilgner A, *et al.* (2021) Acidity and the multiphase chemistry of atmospheric aqueous particles and clouds. *Atmos. Chem. Phys.* 21(17):13483-13536.
7. Pye HOT, *et al.* (2020) The acidity of atmospheric particles and clouds. *Atmos. Chem. Phys.* 20(8):4809-4888.
8. Jang M, Czoschke NM, Lee S, & Kamens RM (2002) Heterogeneous Atmospheric Aerosol Production by Acid-Catalyzed Particle-Phase Reactions. *Science* 298(5594):814-817.
9. Cheng Y, *et al.* (2016) Reactive nitrogen chemistry in aerosol water as a source of sulfate during haze events in China. *Science Advances* 2(12):e1601530.
10. Snider G, *et al.* (2016) Variation in global chemical composition of PM2.5: emerging results from SPARTAN. *Atmos. Chem. Phys.* 16(15):9629-9653.
11. Tao W, *et al.* (2020) Aerosol pH and chemical regimes of sulfate formation in aerosol water during winter haze in the North China Plain. *Atmos. Chem. Phys.* 20(20):11729-11746.
12. Keene WC, *et al.* (1998) Aerosol pH in the marine boundary layer: A review and model evaluation. *Journal of Aerosol Science* 29(3):339-356.
13. Wang W, *et al.* (2021) Sulfate formation is dominated by manganese-catalyzed oxidation of SO2 on aerosol surfaces during haze events. *Nature Communications* 12(1):1993.
14. Liu T, Chan AWH, & Abbatt JPD (2021) Multiphase Oxidation of Sulfur Dioxide in Aerosol Particles: Implications for Sulfate Formation in Polluted Environments. *Environmental Science & Technology* 55(8):4227-4242.
15. Angle KJ, *et al.* (2021) Acidity across the interface from the ocean surface to sea spray aerosol. *Proceedings of the National Academy of Sciences* 118(2):e2018397118.

16. Laskin A, *et al.* (2003) Reactions at Interfaces As a Source of Sulfate Formation in Sea-Salt Particles. *Science* 301(5631):340-344.
17. Alexander B, *et al.* (2005) Sulfate formation in sea-salt aerosols: Constraints from oxygen isotopes. *Journal of Geophysical Research: Atmospheres* 110(D10).
18. Chameides WL & Stelson AW (1992) Aqueous-phase chemical processes in deliquescent sea-salt aerosols: A mechanism that couples the atmospheric cycles of S and sea salt. *Journal of Geophysical Research: Atmospheres* 97(D18):20565-20580.
19. Liao H, Seinfeld JH, Adams PJ, & Mickley LJ (2004) Global radiative forcing of coupled tropospheric ozone and aerosols in a unified general circulation model. *Journal of Geophysical Research: Atmospheres* 109(D16).
20. Salter ME, *et al.* (2016) Calcium enrichment in sea spray aerosol particles. *Geophysical Research Letters* 43(15):8277-8285.
21. Shi G, *et al.* (2017) pH of Aerosols in a Polluted Atmosphere: Source Contributions to Highly Acidic Aerosol. *Environmental Science & Technology* 51(8):4289-4296.
22. Ding J, *et al.* (2019) Aerosol pH and its driving factors in Beijing. *Atmos. Chem. Phys.* 19(12):7939-7954.
23. Fountoukis C & Nenes A (2007) ISORROPIA II: a computationally efficient thermodynamic equilibrium model for $K^+$-$Ca^{2+}$-$Mg^{2+}$-$NH_4^+$-$Na^+$-$SO_4^{2-}$-$NO_3^-$-$Cl^-$-$H_2O$ aerosols. *Atmos. Chem. Phys.* 7(17):4639-4659.
24. Zheng G, Su H, Wang S, Pozzer A, & Cheng Y (2021) Impact of non-ideality on reconstructing spatial and temporal variations of aerosol acidity with multiphase buffer theory. *Atmos. Chem. Phys. Discuss.* 2021:1-29.
25. Alexander B, *et al.* (2005) Sulfate formation in sea-salt aerosols: Constraints from oxygen isotopes. *Journal of Geophysical Research: Atmospheres* 110(D10):D10307.
26. Surratt JD, *et al.* (2007) Effect of Acidity on Secondary Organic Aerosol Formation from Isoprene. *Environmental Science & Technology* 41(15):5363-5369.
27. Hallquist M, *et al.* (2009) The formation, properties and impact of secondary organic aerosol: current and emerging issues. *Atmos. Chem. Phys.* 9(14):5155-5236.
28. Carlton AG, *et al.* (2010) Model Representation of Secondary Organic Aerosol in CMAQv4.7. *Environmental Science & Technology* 44(22):8553-8560.
29. Franco B, *et al.* (2021) Ubiquitous atmospheric production of organic acids mediated by cloud droplets. *Nature* 593(7858):233-237.
30. Gelaro R, *et al.* (2017) The Modern-Era Retrospective Analysis for Research and Applications, Version 2 (MERRA-2). *Journal of Climate* 30(14):5419-5454.

31. Eyring V, Köhler HW, van Aardenne J, & Lauer A (2005) Emissions from international shipping: 1. The last 50 years. *Journal of Geophysical Research: Atmospheres* 110(D17):D17305.
32. Eyring V, Köhler HW, Lauer A, & Lemper B (2005) Emissions from international shipping: 2. Impact of future technologies on scenarios until 2050. *Journal of Geophysical Research: Atmospheres* 110(D17):D17306.
33. Wang C, Corbett JJ, & Firestone J (2008) Improving Spatial Representation of Global Ship Emissions Inventories. *Environmental Science & Technology* 42(1):193-199.
34. Stettler MEJ, Eastham S, & Barrett SRH (2011) Air quality and public health impacts of UK airports. Part I: Emissions. *Atmospheric Environment* 45(31):5415-5424.
35. Stettler MEJ, Boies AM, Petzold A, & Barrett SRH (2013) Global Civil Aviation Black Carbon Emissions. *Environmental Science & Technology* 47(18):10397-10404.
36. Simone NW, Stettler MEJ, & Barrett SRH (2013) Rapid estimation of global civil aviation emissions with uncertainty quantification. *Transportation Research Part D: Transport and Environment* 25:33-41.
37. van der Werf GR, *et al.* (2010) Global fire emissions and the contribution of deforestation, savanna, forest, agricultural, and peat fires (1997–2009). *Atmos. Chem. Phys.* 10(23):11707-11735.
38. Guenther AB, *et al.* (2012) The Model of Emissions of Gases and Aerosols from Nature version 2.1 (MEGAN2.1): an extended and updated framework for modeling biogenic emissions. *Geosci. Model Dev.* 5(6):1471-1492.
39. Hudman RC, *et al.* (2012) Steps towards a mechanistic model of global soil nitric oxide emissions: implementation and space based-constraints. *Atmos. Chem. Phys.* 12(16):7779-7795.
40. Sauvage B, *et al.* (2007) Remote sensed and in situ constraints on processes affecting tropical tropospheric ozone. *Atmos. Chem. Phys.* 7(3):815-838.
41. Carn SA, Yang K, Prata AJ, & Krotkov NA (2015) Extending the long-term record of volcanic SO2 emissions with the Ozone Mapping and Profiler Suite nadir mapper. *Geophysical Research Letters* 42(3):925-932.
42. Shephard MW, *et al.* (2011) TES ammonia retrieval strategy and global observations of the spatial and seasonal variability of ammonia. *Atmos. Chem. Phys.* 11(20):10743-10763.
43. Luo M, *et al.* (2015) Satellite observations of tropospheric ammonia and carbon monoxide: Global distributions, regional correlations and comparisons to model simulations. *Atmospheric Environment* 106:262-277.

44. Whitburn S, *et al.* (2016) A flexible and robust neural network IASI-NH3 retrieval algorithm. *Journal of Geophysical Research: Atmospheres* 121(11):6581-6599.

45. Zeng Y, Tian S, & Pan Y (2018) Revealing the Sources of Atmospheric Ammonia: a Review. *Current Pollution Reports* 4(3):189-197.

46. Luo G, Yu F, & Moch JM (2020) Further improvement of wet process treatments in GEOS-Chem v12.6.0: Impact on global distributions of aerosol precursors and aerosols. *Geosci. Model Dev. Discuss.* 2020:1-39.

47. Sayer AM, Thomas GE, Palmer PI, & Grainger RG (2010) Some implications of sampling choices on comparisons between satellite and model aerosol optical depth fields. *Atmos. Chem. Phys.* 10(22):10705-10716.

48. Luan Y & Jaeglé L (2013) Composite study of aerosol export events from East Asia and North America. *Atmos. Chem. Phys.* 13(3):1221-1242.

49. Wang S, *et al.* (2020) Natural gas shortages during the "coal-to-gas" transition in China have caused a large redistribution of air pollution in winter 2017. *Proceedings of the National Academy of Sciences* 117(49):31018-31025.

50. Weber RJ, Guo H, Russell AG, & Nenes A (2016) High aerosol acidity despite declining atmospheric sulfate concentrations over the past 15 years. *Nature Geosci* 9(4):282-285.

51. Ding AJ, *et al.* (2016) Enhanced haze pollution by black carbon in megacities in China. *Geophysical Research Letters* 43(6):2873-2879.

52. Cheng Y, *et al.* (2016) Reactive nitrogen chemistry in aerosol water as a source of sulfate during haze events in China. *Science Advances* 2(12).

53. Tian S, Pan Y, Liu Z, Wen T, & Wang Y (2014) Size-resolved aerosol chemical analysis of extreme haze pollution events during early 2013 in urban Beijing, China. *Journal of Hazardous Materials* 279:452-460.

54. He H, *et al.* (2014) Mineral dust and NOx promote the conversion of SO2 to sulfate in heavy pollution days. *Sci. Rep.* 4.

55. Wang Y, *et al.* (2014) Mechanism for the formation of the January 2013 heavy haze pollution episode over central and eastern China. *Sci. China Earth Sci.* 57(1):14-25.

56. Behera SN & Sharma M (2010) Investigating the potential role of ammonia in ion chemistry of fine particulate matter formation for an urban environment. *Science of The Total Environment* 408(17):3569-3575.

57. Bencs L, *et al.* (2008) Mass and ionic composition of atmospheric fine particles over Belgium and their relation with gaseous air pollutants. *Journal of Environmental Monitoring* 10(10):1148-1157.

58. Guo H, Weber RJ, & Nenes A (2017) High levels of ammonia do not raise fine particle pH sufficiently to yield nitrogen oxide-dominated sulfate production. *Scientific Reports* 7(1):12109.

59. Alexander B, Park RJ, Jacob DJ, & Gong S (2009) Transition metal-catalyzed oxidation of atmospheric sulfur: Global implications for the sulfur budget. *Journal of Geophysical Research: Atmospheres* 114(D2).

60. Shao J*, et al.* (2019) Heterogeneous sulfate aerosol formation mechanisms during wintertime Chinese haze events: air quality model assessment using observations of sulfate oxygen isotopes in Beijing. *Atmos. Chem. Phys.* 19(9):6107-6123.

61. Zheng B*, et al.* (2015) Heterogeneous chemistry: a mechanism missing in current models to explain secondary inorganic aerosol formation during the January 2013 haze episode in North China. *Atmos. Chem. Phys.* 15(4):2031-2049.